\documentclass[12pt,onecolumn,draftclsnofoot]{IEEEtran}
\usepackage{amsfonts}
\IEEEoverridecommandlockouts
\ifCLASSINFOpdf
\else
\fi
\usepackage{multirow}
\usepackage{booktabs}
\usepackage{makecell}
\usepackage{graphicx}
\usepackage{subfigure}
\usepackage{cite}
\usepackage{color}
\usepackage[pagebackref=false,breaklinks=false,letterpaper=true,colorlinks,bookmarks=false]{hyperref}
\usepackage{amsmath}
\usepackage{amssymb}
\usepackage{array}
\usepackage{bm}
\usepackage{algorithm}
\usepackage{algorithmic}

\usepackage{cite}
\usepackage{lettrine} 

\title{A Novel Automatic Modulation Classification Scheme Based on Multi-Scale Networks}
\author{Hao Zhang, \IEEEmembership{Student Member, IEEE}, Fuhui Zhou, \IEEEmembership{Senior Member, IEEE},\\ Qihui Wu, \IEEEmembership{Senior Member, IEEE}, Wei Wu, and Rose Qingyang Hu, \IEEEmembership{Fellow, IEEE}
\thanks{
H. Zhang, F. Zhou, and Qihui Wu are with College of Electronic and Information Engineering, Nanjing University of Aeronautics and Astronautics, Nanjing 211106 China. They are also with Key Laboratory of Dynamic Cognitive System of Electromagnetic Spectrum Space (Nanjing University of Aeronautics and Astronautics), and with Ministry of Industry and Information Technology, Nanjing, 211106, China (email: haozhangcn@nuaa.edu.cn, zhoufuhui@ieee.org, wuquhui2014@sina.com)

W. Wu is with the Nanjing University of Posts and Telecommunications, Nanjing 210003, China, also with the Key Laboratory of Dynamic Cognitive System of Electromagnetic Spectrum Space, Ministry of Industry and Information Technology, Nanjing University of Aeronautics and Astronautics, Nanjing 211106, China (e-mail: weiwu@njupt.edu.cn).

R. Q. Hu is with the Department of Electrical and Computer Engineering, Utah State University, Logan, UT 84322 USA (e-mail: rosehu@ieee.org).
}
}

\begin{document}

\maketitle

\begin{abstract}
	Automatic modulation classification enables intelligent communications and it is of crucial importance in today's and future wireless communication networks. Although many automatic modulation classification schemes have been proposed, they cannot tackle the intra-class diversity problem caused by the dynamic changes of the wireless communication environment. In order to overcome this problem, inspired by face recognition, a novel automatic modulation classification scheme is proposed by using the multi-scale network in this paper. Moreover, a novel loss function that combines the center loss and the cross entropy loss is exploited to learn both discriminative and separable features in order to further improve the classification performance. Extensive simulation results demonstrate that our proposed automatic modulation classification scheme can achieve better performance than the benchmark schemes in terms of the classification accuracy. The influence of the network parameters and the loss function with the two-stage training strategy on the classification accuracy of our proposed scheme are investigated. 
\end{abstract}


\section{Introduction}
\lettrine[lines=2]{W}{ITH} the commercial applications of the fifth generation (5G) wireless communication networks, the sixth generation (6G) wireless communication networks have received an increasing attention from both academia and industry \cite{saad2019vision} and \cite{letaief2019roadmap}. Intelligent communication empowered by artificial intelligence is one of the most evident characteristics of 6G wireless communication systems. To realize this goal, it is of crucial importance to automatically recognize the modulation types \cite{qin201920}. Automatic modulation classification (AMC)  identifies the modulation type based on the received signals \cite{dobre2007survey}. It is a process needed for decoding the information. Due to its importance, AMC has a wide application in the military and civilian fields, such as spectrum management, electronic warfare, interference identification, and so forth \cite{bisio2018blind} and \cite{qi2020automatic}. 

\subsection{Automatic Modulation Classification Approaches}
The AMC schemes can be classified into two categories, namely, model-driven AMC and data-driven AMC. Specifically, model-driven AMC is to classify the modulation types based on a hypothesis testing or defining specific features while data-driven AMC is to learn features from the data and make the decision. Most of the model-driven classifiers for AMC can be mainly classified into two categories, namely, likelihood-based (LB) classifiers and feature-based (FB) classifiers \cite{dobre2015signal}. The LB classifier treats AMC as a hypothesis testing problem. It calculates the likelihood function under the modulation hypothesis by using the received signal symbols and compares these functions to make the final decision. The FB classifier aims to find better features of the received signals. These two kinds of classifiers have taken the leading positions of AMC over decades since the LB can achieve the optimal solution from the Bayes’ sense while FB can obtain a high reliability for recognizing simple modulation types, such as BPSK and QPSK \cite{hazza2013overview}. However, the LB classifiers have a high computation complexity for estimating unknown parameters \cite{hameed2009likelihood} while the performance of the FB classifier is significantly subjected to the quality of the features. Those disadvantages have limited their further applications into more areas. 

Moreover, model-driven AMC schemes can be only applied to those scenarios where the noise intensity and channel condition are deterministic. However, in practice, noise and channel conditions can be dynamic. Thus, model-driven AMC schemes may not  perform very well \cite{Huang2020}. In order to overcome those problems, data-driven AMC schemes were proposed \cite{dobre2007survey}. The idea of the data-driven schemes is to use features of the received signals to train the classifiers. The machine learning (ML) based classifiers are first trained well with the collected signals under different modulation schemes, and then are used for the modulation classification of the received signal symbols at the test mode. The ML classifiers such as support vector machine (SVM) \cite{anguita2002improved}, K-nearest neighbor (KNN) \cite{samet2007k}, and logistic regression \cite{jiang2018feature} can be readily applied to unknown scenes after being well-trained. However, these ML-based schemes can only classify the features generated by other methods such as feature engineering in the FB schemes, thus the performance is significantly influenced by the feature generation \cite{dobre2007survey}.

With the advancement of computing technologies  and  the integrated circuits, deep learning with deep layers can “learn” features from the original data and automatically perform classifications by using the classifier layers. The past few years have witnessed the unprecedented success of deep learning represented by convolutional neural network (CNN) in the field of computer vision \cite{zhang2019recent} and \cite{xu2020automatic}, nature language processing, \emph{et al}. Compared with the traditional ML-based schemes, deep learning (DL) based schemes, such as CNN \cite{krizhevsky2012imagenet} and recurrent neural network (RNN) \cite{graves2013speech},  incorporate feature extraction and classification as an end-to-end pipeline. For AMC, DL-based schemes can be mainly categorized into two classes, namely, CNN models and RNN models. CNN models identify the signal frames as images and extract features by convolutional operations. RNN and its variants such as long short­ term memory (LSTM) have also been utilized to perform AMC for its ability of capturing “memory” information when dealing with time series data, such as sentences and signals. The details for the related existing research works are presented as follows. 

\subsection{Related Works and Motivation}
O’Shea \emph{et al.} \cite{OShea2016} first exploited a two-layer CNN as the feature extractor and then classified the feature representations with a \emph{softmax} classifier. It is well known that the increase of the depth of the network can boost the feature representation ability \cite{simonyan2014very}, and result in the gradient vanishing and over-fitting problem. In \cite{he2016deep}, the authors designed a deep residual network (ResNet) by using residual learning with skip connection for image classification, which alleviates the over-fitting problem when training deep networks, and is able to learn discriminative features for achieving a better performance. For AMC, the authors in \cite{liu2017deep} directly adopted ResNet to recognize the modulations. However, the classification performance is limited without any modifications of the network structure. O’Shea \emph{et al.} \cite{OShea2018} proposed a modified ResNet by taking real and imaginary (I/Q) parts of the received signals as the input for extracting features. The authors in \cite{huynh2020mcnet} designed convolutional blocks with asymmetric kernels to acquire more powerful features. Instead of concentrating on the CNN structure, Qi \emph{et al.} \cite{qi2020automatic} adopted ResNet to learn multi-modal data combining I/Q data, envelope data and spectrum data. Multiple deep CNNs have been applied for boosting the performance of AMC in \cite{meng2018automatic} and \cite{Wang2020a}. However, a fixed kernel size of $3\times1$ convolution was adopted in these works, which cannot capture the long term spatio-temporal information. 
Meanwhile, West \emph{et al.} \cite{West2017} first proposed a convolutional long short-term deep neural networks (CLDNN) model which combines CNN and LSTM for modulation classification. The direct combination cannot perform very well due to its simple structure. Rajendran \emph{et al.} \cite{rajendran2018deep} and Hu \emph{et al.} \cite{Hu2020} adopted LSTM, which can obtain high accuracy with fewer signal symbols. However, it requires a long training time due to its recurrent structure. Huang \emph{et al.} \cite{Huang2020} utilized an another kind of RNN named gated recurrent unit (GRU) to classify the received signals by exploring the timing correlation characteristics. However, both LSTM and GRU are time-consuming and these models can only be applied to few modulation formats. Typically, most CNNs and RNNs for AMC adopt the \emph{softmax} classifier and the cross entropy loss (simplified as \emph{softmax} loss) to implement the classification. However, the \emph{softmax} loss can only learn separable features.

Besides the basic structures of CNN and RNN for AMC, many other techniques were applied to improve the classification performance, such as feature fusion \cite{zheng2019fusion}, \cite{Zhang2019}, model acceleration\cite{ramjee2019fast}, \cite{wang2020lightamc}, complex networks \cite{tu2020complex}, generative adversarial network (GAN) \cite{patel2020data}, \cite{tu2018semi}, adversarial attacks \cite{lin2020adversarial}. 
Zheng \emph{et. al} \cite{zheng2019fusion} investigated the feature fusion for CNN-base AMC methods, which shown better performance compared to the non-fusion method. 
In \cite{ramjee2019fast}, the authors adopted the principal component analysis to reduce the training time and achieved a good performance under low SNRs. 
Wang \emph{et. al} \cite{wang2020lightamc} designed a lightweight model with smaller model sizes and faster computational speed by using model compression. 
The authors in \cite{tu2020complex} proposed a complex-valued network for AMC which can achieved a superior performance than the real-valued counterparts. 
A conditional generative adversarial network (CGAN) was applied to in \cite{patel2020data} to generate sufficient high-quality labeled data. 
Lin \emph{et. al} \cite{lin2020adversarial} studied the adversarial attacks for improving the reliability and security of AMC. 
These techniques mentioned above can be utilized with the basic structure of CNN or RNN together to obtain a better classification performance.

\begin{figure*}[h]
	\includegraphics[width=0.95\linewidth]{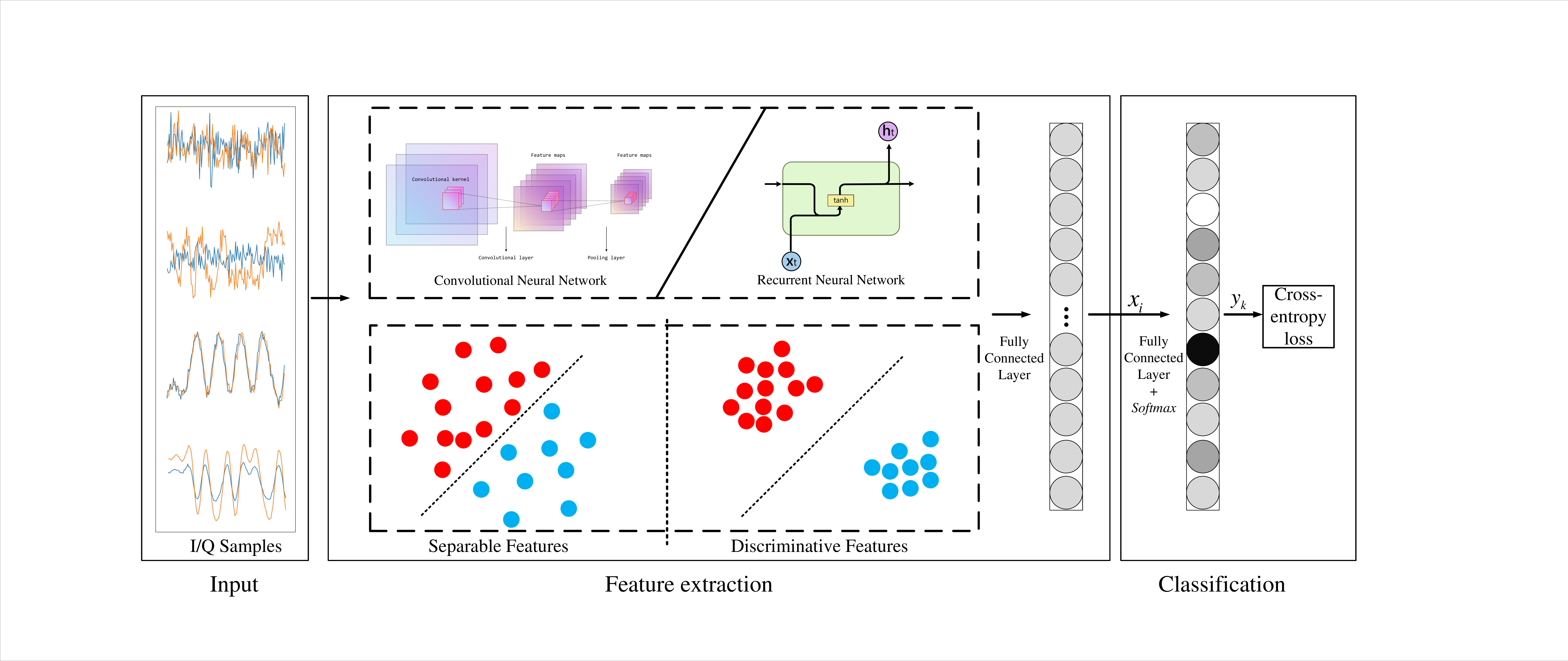}
	\caption{Motivation of the proposed scheme. Intra-class diversity is the major problem that deep CNN/RNN cannot address. These examples are from RadioML 2016.10A\cite{OShea2016}.}
	\label{fig:motivation}
\end{figure*}

The performance of AMC has been greatly improved due to the powerful feature learning ability of  the deep CNN or RNN models. However, the classification performance is subject to the variational noise conditions. As shown in Fig. \ref{fig:motivation}, a typical pipeline of an AMC model consists of several stages, namely, input (data collection of the received signals like I/Q samples), feature extraction by CNN or RNN models, classification by the fully connected (FC) layer with the \emph{softmax} loss. In these CNNs, kernel sizes such as $3\times1$ are usually utilized, similar to the image classification task. However, AMC is different from the image classification. AMC needs to learn a long term spatial correlation while image classification only needs to focus on the neighboring pixels in a small region. Thus, deeper convolutional layers with small kernel sizes cannot capture much spatial information about AMC. RNNs and LSTM can be used for learning long-term correlation. However, they are quite time-consuming. Instead of introducing the recurrent mechanism, CNNs with large kernels can also achieve high performance since larger kernels can learn longer spatial features \cite{he2016deep}. Thus, the utilization of multi-scale convolutions with large kernels can achieve a high classification performance while the computation cost is  maintained reasonably low.

In the DL models, the \emph{softmax} loss is used as the default metric since it is appropriate for classifying features in the space with a high dimension, and it has been applied in the most classification tasks in \cite{he2016deep} and \cite{OShea2018}. It enables the DL models to learn the separable features so that it  can deal with the inter-class diversity \cite{wen2016discriminative}. However, unlike other kinds of data such as images or sentences, which normally have a small difference among the same class, the signals here are  very different under different SNRs due to the existence  of noises. As shown in the input part of Fig. \ref{fig:motivation}, the I/Q samples under the same modulation type with different SNRs ($-6$dB, $0$dB, $6$dB, $12$dB) have diverse wave shapes, which cause challenge for the classifier. The intra-class diversity is the major problem that makes the DL based models not to perform very well under the typical classification pipeline using the \emph{softmax} loss. Under this situation, it is quite important to learn both discriminative and separable features, which are much easier to be classified compared to the separable features. 

The reason why the discriminative features are much easier to be classified is that the distances among the same class features are smaller. That is to say, the discriminative features can decrease the influence of the intra-class diversity caused by different SNRs. However, the existing works that used DL models to perform AMC cannot capture small intra-class scatters. Thus, it is urgent to find a scheme that can learn both separable and discriminative features for AMC.

\subsection{Contributions and Organization}
In this paper, in order to tackle the intra-class diversity problem encountered by the existing AMC schemes based on DL, a novel AMC scheme is proposed by using a novel loss function with a two-stage training strategy and a multi-scale convolutional neural network (MSNet). The novel loss is inspired by the idea from face recognition which learns centering features within mini-batches and the two-stage training strategy can learn both discriminative and separable features. MSNet is constructed by several multi-scale modules, and it is utilized to extract features from different levels. The multi-scale module consisting of several convolutional layers with different kernel sizes is proposed for capturing multi-level features, where large-kernel convolution can learn much more spatio-temporal information. The main contributions of this paper are summarized as follows. 
\begin{enumerate}
	\item 
	The existing AMC schemes based on DL cannot simultaneously utilize discriminative and separable features.
	It is the first time that both the discriminative and separable features are exploited in AMC for tackling the intra-class diversity problem encountered by the existing AMC schemes based on DL. This is achieved by proposing a novel AMC scheme based on a new loss function and a two-stage training strategy. The novel loss is inspired by the idea from face recognition and the two­-stage training strategy can learn both discriminative and separable features. Moreover, the novel loss function and the training strategy can be flexibly integrated into other DL-based schemes.
	\item To learn multi-level features with spatio-temporal information for obtaining more separable features, MSNet with multi-scale convolutional layers and large kernels is proposed for AMC. The multi-scale module is designed to learn deep features from high dimensional tensors by using several parallel convolutional layers with different kernel sizes. Moreover, a large­-kernel convolution is utilized to extract long correlation of the spatial information. To enable the MSNet to take samples with an arbitrary length, global averaging pooling is used to generate features with a fixed length for later FC layers.
	\item Extensive simulation results demonstrate the superiority of our proposed AMC scheme compared with other DL-based AMC schemes in terms of the classification accuracy. Moreover, the visualized features present the discriminative and separable representation ability and the superiority of our proposed AMC scheme from another aspect. In addition, the influence of the network parameters and the loss function with the two-stage training strategy on performance of our proposed scheme is investigated. 
\end{enumerate}

The rest of this paper is organized as follows. Section II presents the signal model and the DL-based AMC schemes. Section III presents our proposed AMC scheme. Simulation results are given in Section IV and Section V concludes the paper. 

\section{Problem Formulation}
Generally, modulation classification can be identified as a $K$-class hypothesis test, where $K$ denotes the class number of the  modulations. According to the traditional modulation classification models \cite{dobre2007survey}, where the receiver is equipped with a single antenna, the received signal under $k$th modulation hypothesis $H_k$ is given by
\begin{equation}
H_k: x_k(n)=s_k(n)+\omega_k(n), n=1,2,...,N,
\end{equation} 
where $s_k(n)$ and $x_k(n)$ represent the transmitted signal and the received signal respectively, $N$ is the number of signal symbols, and $\omega_k(n)$ denotes the additive white Gaussian noise (AWGN) with zero mean and variance $\delta^2_\omega$.

In this paper, the real and imaginary parts of the received signal from the In-phase and Quadrature (I/Q) parts are both utilized. These two parts usually obey an identical independent distribution for a signal sample, which can be inputs into the neural network without normalization \cite{OShea2018}. It is worthy noting that the combination of those two parts of I/Q samples carries the universal features of the most modulations. Other types of signals such as cyclic spectrum \cite{Li2018}, constellation diagrams \cite{lin2020contour}, SPWVD and BJD \cite{Zhang2019}, and high-order cumulants \cite{9042348}, can also be used for AMC. The I/Q signal samples can be expressed as a vector by turning $x_k(n)$ into the vector $\textbf{\text{x}}_k$, given as
\begin{align}
\textbf{\text{x}}_k &=\textbf{\text{I}}_k+\textbf{\text{Q}}_k\nonumber\\
 &=\Re{(\textbf{\text{x}}_k)}+j\Im{(\textbf{\text{x}}_k)},
\end{align}
where $\textbf{\text{I}}_k$ and $\textbf{\text{Q}}_k$ represent the real and imaginary parts of the signals, respectively, and $j=\sqrt{-1}$. $\Re(\cdot)$ and $\Im(\cdot)$ represent the operators of the real and imaginary parts of the signal, respectively. $\textbf{\text{x}}_k$ is specifically expressed as
\begin{equation}
\textbf{\text{x}}^{IQ}_k = \left(
\begin{array}{c}
\Re[x(1), x(2), \ldots, x(N)]\\
\Im[x(1), x(2), \ldots, x(N)]\\
\end{array} \right).
\end{equation}

As shown in Fig. \ref{fig:motivation}, a large number of the received signal samples are collected and the DL-based schemes (CNN or RNN) are utilized to extract features from these raw data. Then, the extracted features are further input into the fully connected layers for integrating the information. CNN and RNN such as ResNet\cite{OShea2018} or LSTM\cite{rajendran2018deep} are adopted to learn spatial and temporal features of the I/Q samples, and the classification is operated by using these learned features. The feature learning can be expressed as a process that the raw data $\textbf{\text{x}}_k\in\mathbb{R}^{N\times 2}$ is mapped into a $L$-dimensional vector $\bm{x}_i$, given as
\begin{equation}
f: \textbf{\text{x}}_k\in\mathbb{R}^{N\times 2}\to
\bm{x}_i\in\mathbb{R}^{L},
\end{equation}
where the mapping function $f$ represents the CNN or the RNN model with the FC layer, and $\bm{x}_i$ represents the features from the $i$th layer, $\bm{x}_0=input(\textbf{\text{x}}_k)$ represents the \emph{tensor} conversion from the raw input.

Then, the learned features are classified by the last FC layer, which is the combination of the FC layer and the \emph{softmax} classifier. The number of neurons in the last layer is equal to the number of modulation formats. Thus, each output neuron corresponds to a modulation format. \emph{Softmax} is utilized to convert the output into the probability with which the input signal belongs to each candidate modulation format. \emph{Softmax} activation function can be expressed as
\begin{equation}
y_k=\frac{\exp(\bm{w}^T_{y_k}\bm{x}_i+b_{y_c})}{\sum^K_{j=1}\exp(\bm{w}^T_{j}\bm{x}_i+b_{j})},
\end{equation}
where $y_k$ is the output of the $k$th neuron (i.e. the probability of belonging to the $k$th class). $\bm{x}_i\in\mathbb{R}^d$ denotes the $i$th deep feature, and $d$ is the feature dimension. $\bm{w}_j\in\mathbb{R}^d$ represents the $j$th column of the weight matrix $\bm{W}\in\mathbb{R}^{d\times K}$ in the last fully connected layer and $\bm{b}\in\mathbb{R}^d$ is the bias term. $K$ is the number of classes. The cross-entropy loss function is utilized to learn the correct label prediction, given as
\begin{equation}
\mathcal{L}_{S}=-\sum^{m}_{i=1}\log y_i,
\label{cross}
\end{equation} 
where the loss function can also be formulated as $\mathcal{E}(f,\bm{\theta})$, and $\bm{\theta}$ are the parameters required to be optimized. 
Suppose that the received signals $x_0$ are generated from the $k$th modulation $\mathcal{M}_k$. 
Then replacing $y_i$ with the physical meaning, $\Pr(\mathcal{M}_k;f,\bm{\theta|x})$, the loss function can be expressed as 
\begin{align}
\mathcal{E}(f,\bm{\theta})=-\sum^{m}_{i=1}\log \Pr(\mathcal{M}_i;f,\bm{\theta|x})\nonumber\\
=-\log \Pr(\mathcal{M}_j;f,\bm{\theta|x}),
\label{ef}
\end{align} 
where $\mathcal{M}_j$ is the true modulation scheme and the minimization problem can be formulated as the maximization of $\Pr(\mathcal{M}_j;f,\bm{\theta|x})$, which is equal to the the maximum \emph{aposteriori} probability (MAP) criterion. Thus, the modified MAP criterion can be expressed as
\begin{equation}
\hat{\mathcal{M}}_{i}=\arg\max_{\mathcal{M}_i}\Pr(\mathcal{M}_j;f,\bm{\theta|x}).
\label{cross}
\end{equation}

The \emph{softmax} classifier with the cross-entropy loss is a typical combination for the classification tasks. From eq. (\ref{cross}), only the probability of the correctly categorized class is calculated during the training thus   the differences of the mis-categorized classes are ignored. In other words, the \emph{softmax} loss only concentrates on learning the inter-class diversity, which can make the learned features separable as shown in Fig. \ref{fig:motivation}. However, for AMC, it is difficult to classify the modulation types of the signal samples under various SNRs, which means that the model should also consider the intra-class diversity. The learned features by the \emph{softmax} loss are not discriminative since they still have significant intra-class variations\cite{wen2016discriminative}. Thus, a new loss function which can learn both separable and discriminative features is required for AMC.

\section{Proposed AMC Scheme}
\subsection{Framework of The Proposed Scheme}
\begin{figure*}
	\includegraphics[width=0.95\linewidth]{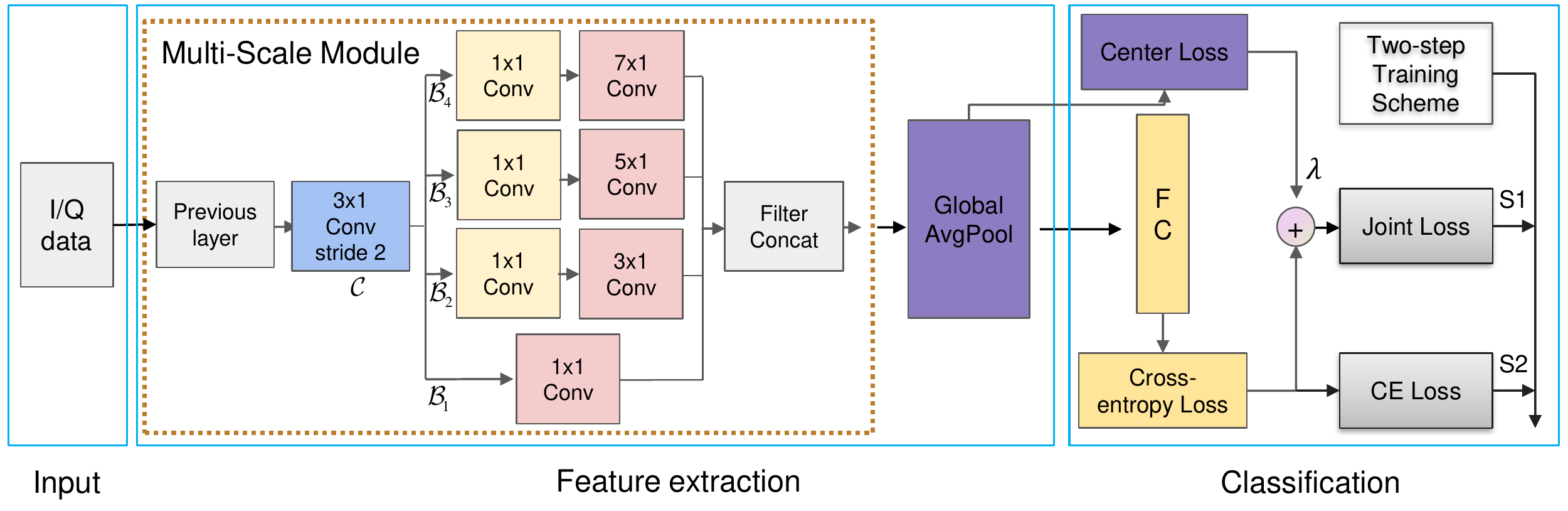}
	\caption{The framework of the proposed scheme.
}
	\label{fig:model}
\end{figure*}

Fig. \ref{fig:model} illustrates the core idea of the proposed scheme. Similar to the typical process of the DL-based models, the proposed scheme consists of three stages, namely, input stage, feature extraction stage, and classification stage. Firstly, in the input stage, the raw data of the signal samples are converted into \emph{tensors} to match the requirements of the deep learning framework, such as PyTorch\cite{paszke2019pytorch}. Then, a multi-scale convolutional network (MSNet) is used to extract features from the input signals. In MSNet, the multi-scale module is powerful for capturing multi-level spatial information, and the global averaging pooling is utilized to adapt for the input length. Moreover, the convolutions with larger kernel sizes can extract more separable features than those with the  smaller sizes. Finally, in the classification stage, besides minimizing the cross-entropy loss (i.e., the \emph{softmax} loss from the final FC layer used for the conventional DL models), a center loss term on the features is introduced for enforcing the models to be more discriminative. Furthermore, a two-stage training strategy is utilized to make the MSNet more discriminative and separable. The MSNet is first trained with the joint supervision of center loss and \emph{softmax} loss to learn discriminative features for AMC. Then, it can be further optimized by the \emph{softmax} loss. The center loss and the training strategy can be used in other state-of-the-art DL schemes for extracting more powerful features to further improve the accuracy.

\subsection{Multi-Scale Convolutional Neural Network}
In this paper, a MSNet is designed to learn more separable features for improving the classification accuracy of AMC. As shown in Fig. \ref{fig:model}, the proposed MSNet consists of several multi-scale modules, global average-pooling (GAP), FC layers and the \emph{softmax} classifier. Multi-scale modules are used to capture the  multi-level feature information by using a convolution layer with $stride=2$ to reduce the feature dimension at the top layer of the module and several parallel convolutions with different kernel sizes. Then, GAP is applied for aggregating information from the convolutional layers. Moreover, GAP enables the whole network to take samples with an arbitrary length. Lastly, two FC layers are used to reduce the feature dimension and generate classification results (the last FC layer is with \emph{softmax} classifier).

As the core of the MSNet, the multi-scale (MS) module is illustrated in the left part of Fig. \ref{fig:model}. The MS module is constructed by convolution operations with different kernel sizes within the same layer to capture different levels of features. Firstly, a $3\times 1$ convolutional layer $\mathcal{C}$ with $stride=2$ is utilized to reduce the feature dimension without damaging the spatial information caused by pooling operation. Then, four branches of convolutional layers $\mathcal{B}_i, i=\{1,2,3,4\}$ with different kernel sizes are organized in parallel to extract features with multi-scale information. The kernel sizes of these four branches are $\mathcal{B}_1=1\times 1$, $\mathcal{B}_2=3\times 1$, $\mathcal{B}_3=5\times 1$, and $\mathcal{B}_4=7\times 1$, respectively, with an additional $1\times 1$ convolutional layer to gather information. These branches of convolutional layers increase the width of the network and enable the model  to learn the multi-scale information. Features from different convolutional layers are consolidated together by using $concat$ operation in channel-wise \cite{huynh2020mcnet}, which  increases the non-linear capacity of the trainable network. The MS module can be represented as
\begin{equation}
	\bm{x}_{i+1}=concat(\mathcal{B}_1(\mathcal{C}(\bm{x}_i)), \mathcal{B}_2(\mathcal{C}(\bm{x}_i)), \mathcal{B}_3(\mathcal{C}(\bm{x}_i)), \mathcal{B}_4(\mathcal{C}(\bm{x}_i))),
\end{equation}
where $\bm{x}_i$ represents the feature maps from the previous layer, and $\bm{x}_{i+1}$ denotes the output of the module. The application of the MS module is inspired by the Inception module in GoogLeNet\cite{szegedy2015going} by adding an additional larger kernel $7\times 1$ to capture much more spatial information within one layer. Theoretically, a large kernel is equivalent to the stacking of several small kernels. For example, a $7\times1$ kernel can be represented as three layers of $3\times1$ kernels, i.e. $\mathcal{B}_4=3\mathcal{B}_2$\cite{szegedy2016rethinking}. However, the deep layers with small kernels cannot learn multi-scale information. Additionally, max-pooling operation is removed and replaced by the above-mentioned convolution and batch normalization (BN) is utilized to normalize activations in the convolutional layers. The feature maps generated from these layers form the output feature matrix, which is given as $\bm{x}_i^G=g(\bm{x}_0|\bm{\theta}_0)\in\mathbb{R}^{C\times H\times W}$.  $\bm{x}_0$ is the input signals, $\bm{\theta}_0$ denotes the weights of the MS modules, $C$ is the channel number, and $H\times W$ is the shape of the feature maps. 

\begin{figure}
	\includegraphics[width=0.95\linewidth]{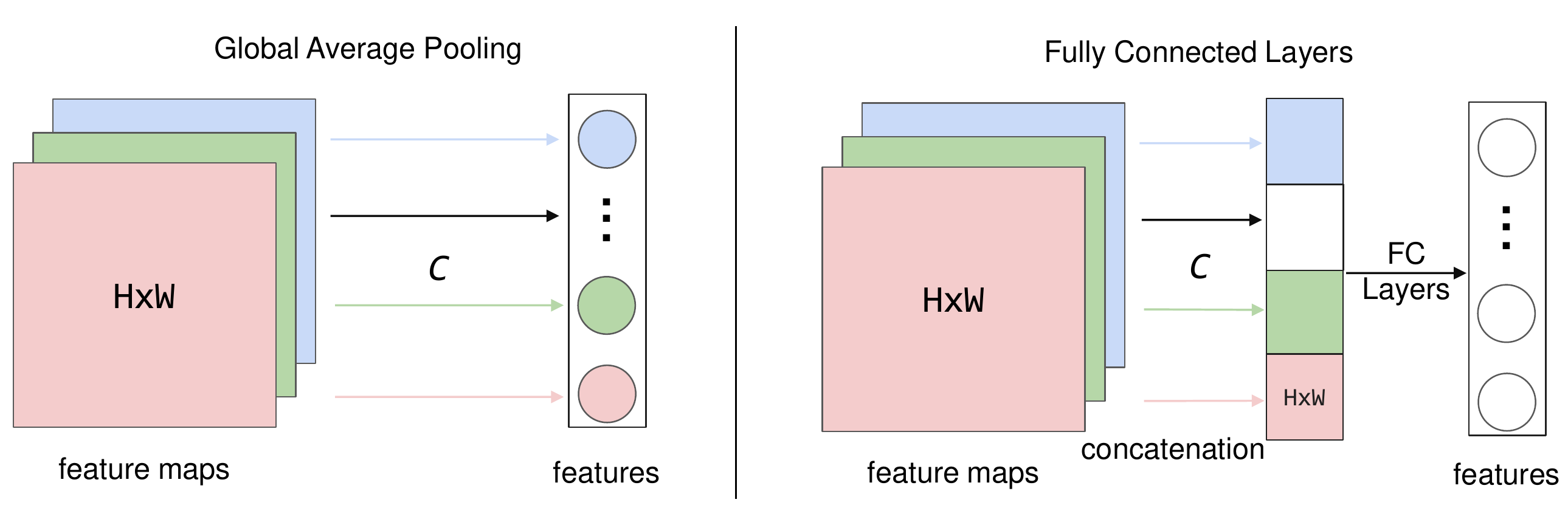}
	\caption{Comparison of global average pooling (GAP) and fully connected (FC) layers.
}
	\label{fig:gap}
\end{figure}
GAP is adopted as a pooling operation and FC layers are utilized to reduce the feature dimension and produce the classification results. As shown in Fig. \ref{fig:gap}, the feature maps generated by the MS modules are the input into the GAP to average each feature map for each corresponding channel in the last convolutional layer. In the original FC layer, the feature maps from convolutional layers are flattened and permuted to a $1$-dimensional vector with a length of $C\times H\times W$, and then the FC layer with $L$ neurons is used to downscale the dimension of these features. Thus, the number of the connections between the convolutional layer and the FC layer is $L\times C\times H\times W$, which leads to  costly computation. GAP is used before the FC layer to enable the arbitrary input length of the signals. One advantage of GAP is that it is more native for the convolution structure by averaging the spatial information within the whole feature map. Another advantage is that there is no parameter required to be optimized in GAP. Thus, over-fitting problem is avoided. Moreover, the output dimension of GAP is $C\times 1\times 1$, and the computation of the later FC layer is reduced to $L\times C$. The features generated by the middle FC layer are  given as 
\begin{align}
\bm{x}_i^{FC} &=fc(\bm{x}_i^G|\bm{\theta}_1)\nonumber\\
 &=fc(g(\bm{x}_0|\bm{\theta}_0)|\bm{\theta}_1)\nonumber\\
 &=f(\bm{x}_0|\bm{\theta})\in\mathbb{R}^{L},
\end{align}
where $\bm{\theta}_1$ denotes the weights of the FC layer, and $\bm{\theta}$ represents the weights of the MSNet. Rectified linear units (ReLU) is used as the activation function to introduce non-linearity into the network. Moreover, ReLU is beneficial for avoiding gradient vanishing and explosion and it also increases the sparsity of the network and alleviates the over-fitting problem. ReLU function can be expressed as
\begin{equation}
f(z)=\max(0,z).
\end{equation}

Finally, an FC layer with \emph{softmax} classifier is served as the classification module. The feature $\bm{x}_i^{FC}$ from the previous feature extraction module is the input into the \emph{softmax} layer with $K$ neurons, and thus, each output corresponds to a modulation type. In the typical DL-based schemes for AMC, the output of the \emph{softmax} layer is used for the cross-entropy loss. In this paper, the feature from the feature extraction module and the output of the classification module are utilized. The hyper-parameters of our proposed MSNet are shown in Table \ref{tab2}, and each convolutional layer consists of $32$ hidden units. In the traditional CNN models, the size of the input data should be fixed to match the FC layer. However, GAP can generate features into a fixed dimension and guarantee that MSNet can process data with any arbitrary length. 

\begin{table}[h]
\centering
\caption{Hyper-parameters of the MSNet}
\begin{center}
\begin{tabular}{ccc}
\toprule
Layer &  Kernel Size &Output \\
\midrule
input & - &$2\times 128$\\
MS module & $32@(3\times 1),[32@\mathcal{B}_1,32@\mathcal{B}_2,32@\mathcal{B}_3,32\mathcal{B}_4] $  &  $128\times 64$\\
MS module & $32@(3\times 1),[32@\mathcal{B}_1,32@\mathcal{B}_2,32@\mathcal{B}_3,32\mathcal{B}_4] $  &  $128\times 32$\\
GAP & $(1\times 1)$ &  $128\times 1$\\
FC & $128$ & $128$\\
FC/\emph{softmax} & $8$ & $8$ \\
\bottomrule
\end{tabular}
\label{tab2}
\end{center}
\end{table}

\subsection{Learning Discriminative Models}

\begin{figure}[h]
	\includegraphics[width=0.95\linewidth]{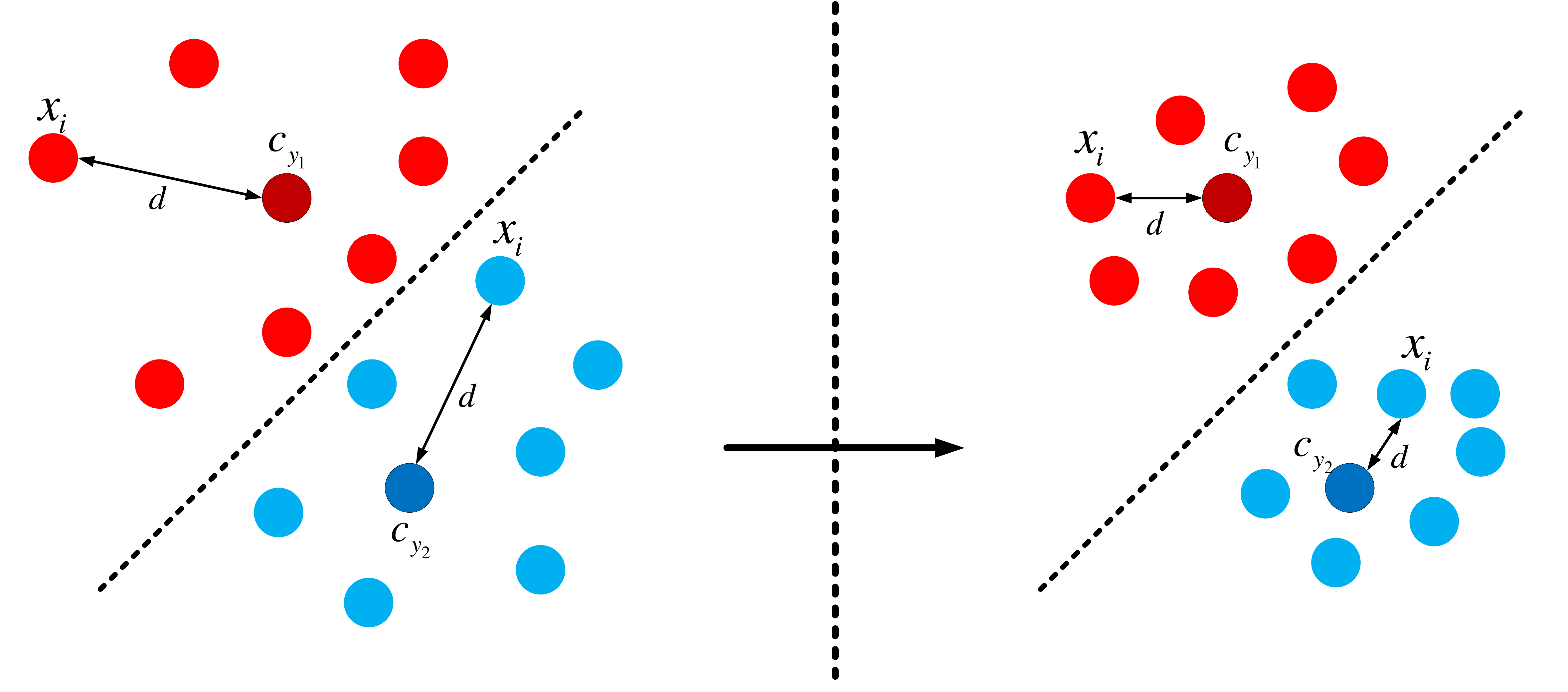}
	\caption{Illustration of the process of learning discriminative features.
}
	\label{fig:center}
\end{figure}

MSNet with multi-scale convolutional layers can learn separable features since it alleviates the inter-class variations. However, the learned features are not discriminative enough due to the intra-class diversity. As shown in the right part of Fig. \ref{fig:model}, an additional term called center loss is introduced for learning discriminative features and the loss alleviates the intra-class variations by minimizing the distances among the learned features. Moreover, a two-stage training strategy is utilized to learn both separable and discriminative features for obtaining a better classification performance.

As shown in Fig. \ref{fig:center}, to overcome the intra-class diversity, the distances $d$ among the features within the same class should be small. 
For this purpose, inspired by face recognition\cite{wen2016discriminative}, 
the concept of center features is introduced for AMC. 
For each modulation, a center feature $\bm{c}_{y_k}$ is assumed, and $d=||\bm{x}_i-\bm{c}_{y_k}||^2_2$ is used to measure the distance between the feature $\bm{x}_i$ and the center feature. 
The center loss function is given as 
\begin{equation}
\mathcal{L}_C=\frac{1}{2}\sum^m_{i=1}||\bm{x}_i-\bm{c}_{y_k}||^2_2,
\end{equation}
where $m$ represents the mini-batch number, $\bm{c}_{y_k}\in\mathbb{R}^d$ denotes the $k$th class center of  the deep features, and $\bm{x}_i$ represents the deep features that are derived from the middle FC layer before the last classification layer. The formulation measures the distance between any two features and also finds the center feature for each class. It  effectively characterizes the intra-class variations\cite{wen2016discriminative}. The center features $\bm{c}_{y_k}$ should be updated when the features are changed. Thus, it is required to find the center features for each class in the entire training set, which is inefficient and even impractical in reality. Instead of finding the center features in the whole training dataset, we iterate the progress based on the mini-batch during the training process. In each iteration, the center features are computed by averaging the features from the corresponding classes (some centers may not be updated). Moreover, to avoid large perturbations caused by few mislabeled samples, a scalar $\alpha\in[0,1]$ is used to control the learning rate of the centers. Then, the gradients of $\mathcal{L}_C$ with respect to $\bm{x}_i$ and the update equation $\bm{c}_{y_k}$ are computed as
\begin{equation}
\frac{\partial\mathcal{L}_C}{\partial \bm{x}_i}=\bm{x}_i-\bm{c}_{y_k},
\end{equation}
\begin{equation}
\Delta \bm{c}_j=\frac{\sum^{m}_{i=1}\delta(y_c=j)(\bm{c}_{y_k}-\bm{x}_i)}
{1+\sum^{m}_{k=1}\delta(y_k=j)},
\end{equation}
where $\delta(condition)=1$ if the $condition$ is satisfied; otherwise $\delta(condition)=0$. The joint supervision of the center loss and \emph{softmax} loss for training the DL models in order to obtain the discriminative feature learning is formulated as
\begin{equation}
\begin{aligned}
\mathcal{L} &=\mathcal{L}_S+\mathcal{L}_C\\
 &=-\sum^{m}_{i=1}\log\frac{\exp({\bm{w}^T_{y_k}\bm{x}_i}+b_{y_k})}{\sum^{K}_{j=1}\exp({\bm{w}^T_{j}\bm{x}_j}+b_{j})}+\frac{\lambda}{2}\sum^{m}_{i=1}||\bm{x}_i-\bm{c}_{y_k}||^2_2,
\end{aligned}
\end{equation}
where a scalar $\lambda$ is used for balancing the two loss functions. The model that jointly supervises the center loss and \emph{softmax} loss can be trained and optimized by using the standard stochastic gradient descent (SGD) algorithm\cite{wen2016discriminative}.

In the existing studies contrastive loss \cite{sun2014deep} and triplet loss \cite{schroff2015facenet} have been proposed to enhance the ability of learning discriminative features. For AMC, the contrastive loss was introduced in \cite{huang2019automatic} for obtaining the discriminative representations. However, the contrastive loss and triplet loss are pair-based methods. They need to find sample pairs or sample triplets from the training set. Thus the efficiency goes down when the size of the training data  increases. As a comparison, in our proposed scheme, the center loss has the same requirement as the \emph{softmax} loss and it does not need complex sample pairs or sample triplets. Consequently, the supervised learning of our MSNet is more efficient and easier to implement. Moreover, the \emph{softmax} loss function directly focuses on the learning objective of the intra-class diversity, which is more efficient for learning the discriminative features.

\subsection{Two-Stage Training Strategy}
\begin{figure*}[h]
	\includegraphics[width=0.95\linewidth]{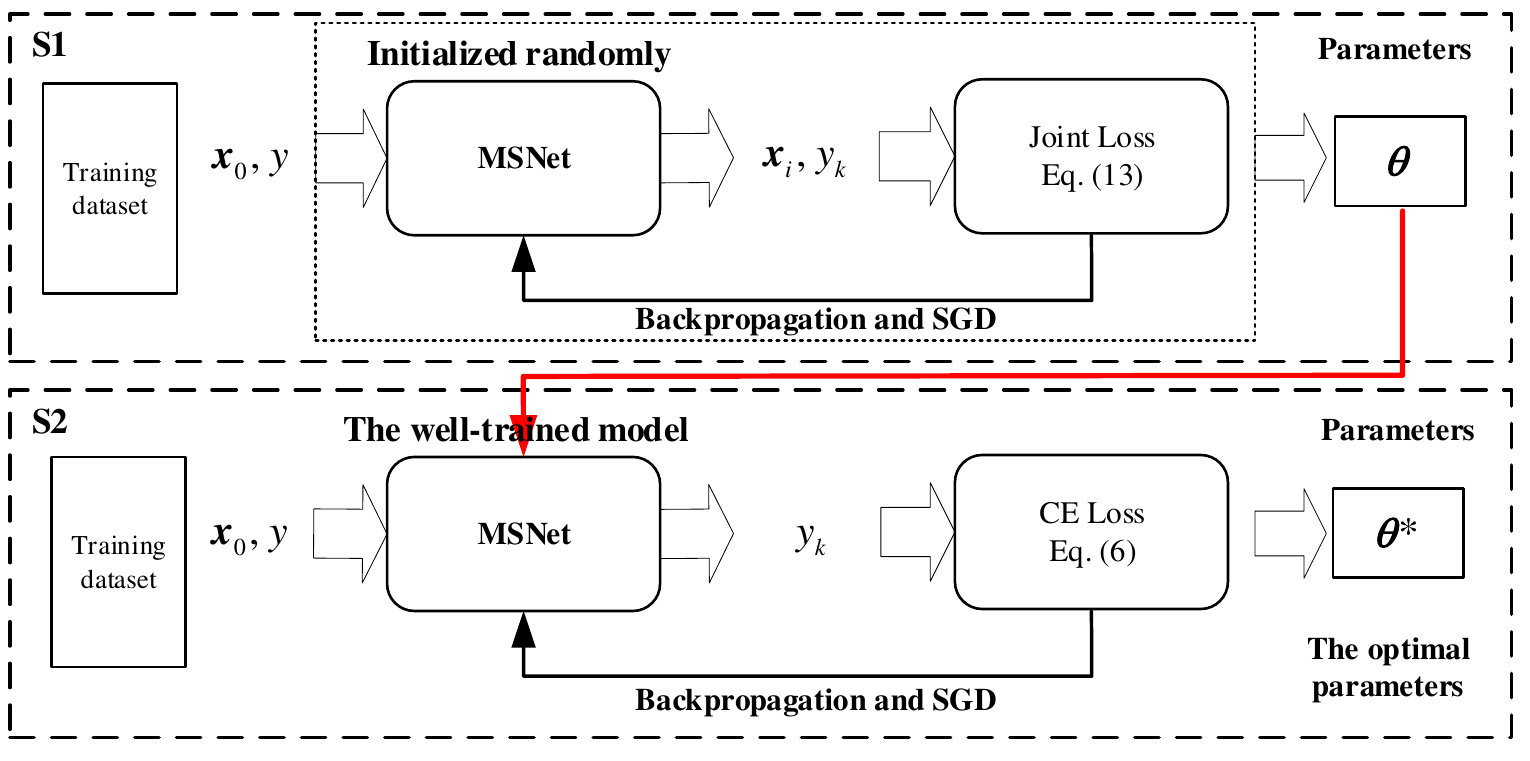}
	\caption{The procedure of the two-stage training strategy.
}
	\label{fig:train}
\end{figure*}

In a general training process, the DL models are trained epoch by epoch by using the \emph{softmax} loss. In each epoch, all the training samples are utilized once to train the model. Within an epoch, the entire training set is shuffled randomly and split into batches with size $m$. Then the model is trained batch by batch. To learn the discriminative features while keeping them separable, a two-step training scheme is proposed in this paper. 

As shown in Fig. \ref{fig:train}, the two-stage training scheme consists of two steps \textbf{S1} and \textbf{S2}, and the same training dataset is used for the training in both steps. In the training dataset, $\bm{x}_0$ and $\bm{y}$ represent the input data and corresponding class label, respectively. In the first step \textbf{S1}, the model is initialized randomly and trained under the joint supervision of center loss and \emph{softmax} loss by using back-propagation and SGD. The training is complete when it converges, and the optimal $\bm{\theta}$ is obtained. 

Then, in the second step \textbf{S2}, the model is initialized from the well-trained model achieved by step \textbf{S1}. Only the \emph{softmax} loss is utilized for further improving the performance of the DL models. When the model converges, the further optimized $\bm{\theta}^*$ is achieved. After the two-step training, the model is able to learn  both the discriminative and separable features. Algorithm \ref{alg:training} summarizes the two-step learning details in the DL models with the joint supervision and the \emph{softmax} loss. One advantage of the two-step training is that the first step \textbf{S1} enables the network to learn both separable and discriminative features. Another advantage is that the second step training can be used to further improve the performance of AMC, which is similar to the fine-tuning in \cite{meng2018automatic}.

\begin{algorithm} 
\caption{The Two-Stage Training Algorithm.} 
\label{alg:training} 
\begin{algorithmic}[1] 
\REQUIRE Training data $\{\bm{x}_0\}$. Initialized parameters $\bm{\theta}$ in MSNet, parameters $\bm{W}$ and $\{\bm{c}_j|j=1,2,\ldots, n\}$ in loss layers, respectively. Initialized hyper-parameter $\lambda$, learning rate $\mu_t$, and the number of iteration $t=0$;
\ENSURE The parameters $\bm{\theta}$.
\STATE Begin the training stage \textbf{S1}.
\STATE \textbf{while} not convergence \textbf{do}
\STATE \quad $t=t+1$;
\STATE \quad Compute the joint loss by $\mathcal{L}^t=\mathcal{L}^t_S+\mathcal{L}^t_C$;
\STATE \quad Compute the back-propagation error $\frac{\partial\mathcal{L}^t}{\partial\bm{x}^t_i}$ for each $i$ by $\frac{\partial\mathcal{L}^t}{\partial\bm{x}^t_i}=\frac{\partial\mathcal{L}^t_S}{\partial\bm{x}^t_i}+\lambda\frac{\partial\mathcal{L}^t_C}{\partial\bm{x}^t_i}$;
\STATE \quad Update the parameters $\bm{W}$ by $\bm{W}^{t+1}=\bm{W}^T-\mu^t\frac{\partial\mathcal{L}^t}{\partial \bm{W}^t}=\bm{W}^T-\mu^t\frac{\partial\mathcal{L}^t_S}{\partial \bm{W}^t}$;
\STATE \quad Update the parameters $\bm{c}_j$ for each $j$ by $\bm{c}^{t+1}_{j}=\bm{c}^{t}_{j}-\alpha\Delta\bm{c}^{t}_{j}$;
\STATE \quad Update the parameters $\bm{\theta}$ by $\bm{\theta}^{t+1}=\bm{\theta}^{t}-\mu^t\sum^m_i\frac{\partial\mathcal{L}^t}{\partial\bm{x}^t_i}\frac{\partial\bm{x}^t_i}{\partial\bm{\theta}^{t}}$;
\STATE \textbf{end while}
\STATE Begin the training stage \textbf{S2}.
\STATE Initialized the parameters $\bm{\theta}$, $\bm{W}$ from the converged model in \textbf{S1}.
\STATE \textbf{while} not convergence \textbf{do}
\STATE \quad $t=t+1$;
\STATE \quad Compute the loss by $\mathcal{L}^t=\mathcal{L}^t_S$;
\STATE \quad Compute the back-propagation error $\frac{\partial\mathcal{L}^t}{\partial\bm{x}^t_i}$ for each $i$ by $\frac{\partial\mathcal{L}^t}{\partial\bm{x}^t_i}=\frac{\partial\mathcal{L}^t_S}{\partial\bm{x}^t_i}$;
\STATE \quad Update the parameters $\bm{W}$ and $\bm{\theta}^*$ following step 6 and 8;
\STATE \textbf{end while}
\end{algorithmic}
\end{algorithm}

\section{Numerical Results and Discussion}
In this section, numerical results are presented to evaluate the proposed scheme operated on the publicly available dataset RadioML 2016.10A \cite{OShea2016}, which is presented on the website \footnote{https://www.deepsig.ai/datasets}. The dataset is generated by the GNU Radio consisting of $11$ modulations ($8$ digital and $3$ analog) under an SNR range from $-20$dB to $20$dB with a $2$dB step. There are $1000$ samples with $128$ points for each modulation type under different SNRs. Thus, the whole dataset consists of a total number of $220,000$ I/Q vectors.

Similar to the work in \cite{Huang2020}, the digital modulations $\phi_C=$\{8PSK, BPSK, CPFSK, GFSK, PAM4, QAM16, QAM64, QPSK\} are selected with an appropriate range of SNRs from $-6$dB to $14$dB with a step of $2$dB to evaluate the models. A total number of $88,000$ vectors are utilized, in which $80$\% of them are used for training and the rest $20$\% are utilized for testing. 

For our proposed MSNet, we train the model by using SGD with an initial learning rate of $0.01$, a weight decay of $5\times 10^{-4}$, and  a momentum of $0.9$ for $40$ epochs. 
When the model is trained with the center loss, different initial learning rates are applied for the center loss and the \emph{softmax} loss, where the center loss adopts an initial learning rate of $1\times 10^{-4}$. 
Then, in the second training stage, the hyper-parameters remain the same and the model can reach a best performance with only several epochs (within $5$ epochs).

To further demonstrate the effectiveness of our proposed model and the training scheme, a large and complicated dataset RadioML 2018.01A \cite{OShea2018} is utilized. 
The RadioML 2018.01A contains 24 kinds of modulations under an SNR range from $-20$dB to $30$dB with a step of $2$dB. 
There are over 2 millions of samples with 1024 points in this dataset. 
For the complicated dataset, we follow the training strategy of the RadioML 2016.10A, and the number of training epoch is increased to $100$. 
The simulation results for the network parameters, the training loss, the feature visualization are operated on the selected RadioML 2016.10A, and the simulation results under the RadioML 2018.01A are also presented. 

To compare the simulation results of the proposed scheme with other models, we re-implement these models and carry out two strategies to guarantee their performances. 
On one hand, we calculate the numbers of model parameters and compare them with the numbers given in \cite{OShea2018}. On the other hand, we train the reconstructed models and evaluate them in the public datasets, and we compare the results obtained in our paper with those presented in \cite{OShea2018,huynh2020mcnet,ramjee2019fast}. 
Thus, we can make sure that the reconstructed model can achieve the actual performances from those two perspectives (model and performance).

\begin{figure*}
\centering
\subfigure[]{
\begin{minipage}[t]{0.45\linewidth}
\centering
	\includegraphics[width=0.95\linewidth]{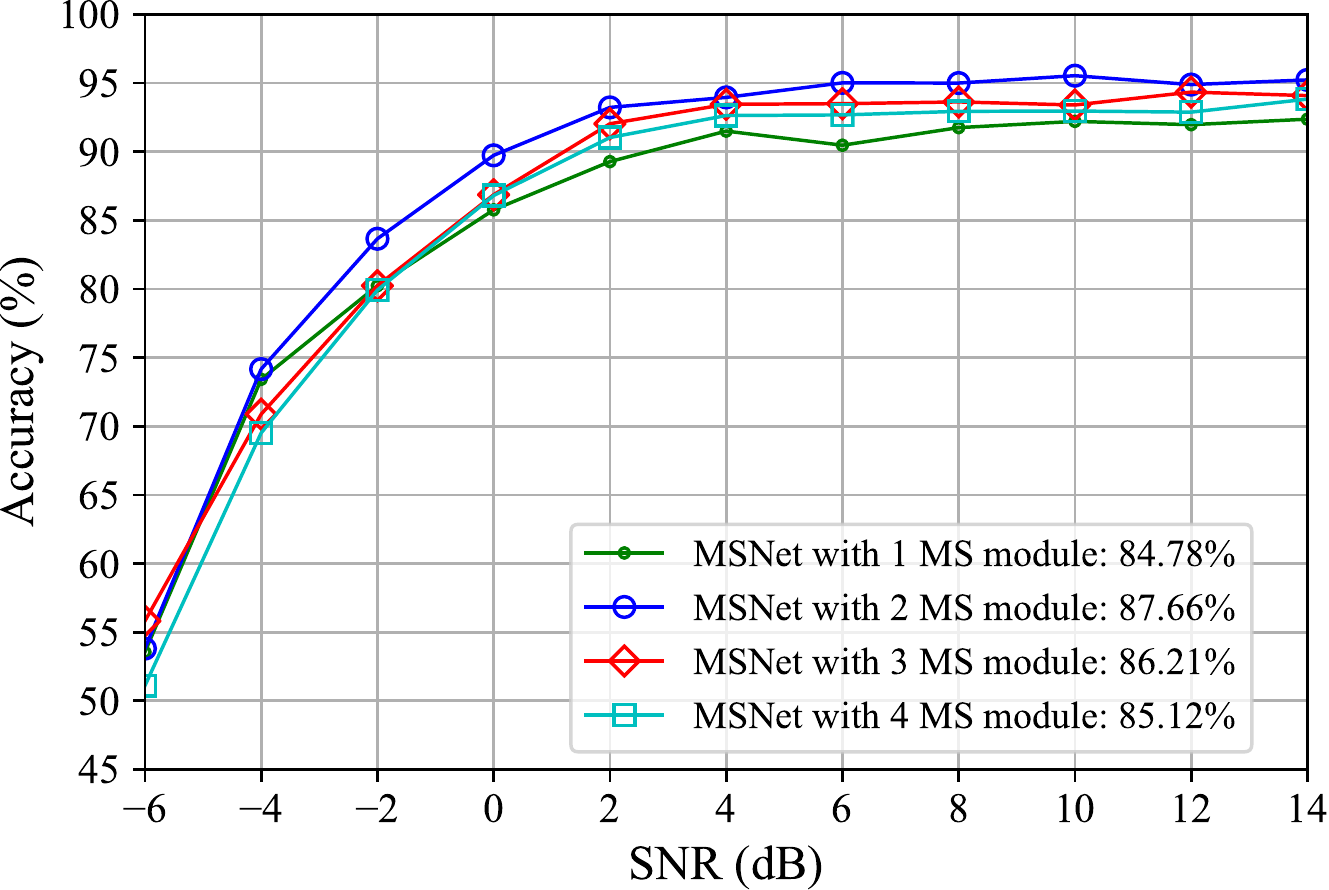}
\label{fig:depth}
\end{minipage}
}
\subfigure[]{
\begin{minipage}[t]{0.45\linewidth}
\centering
	\includegraphics[width=0.95\linewidth]{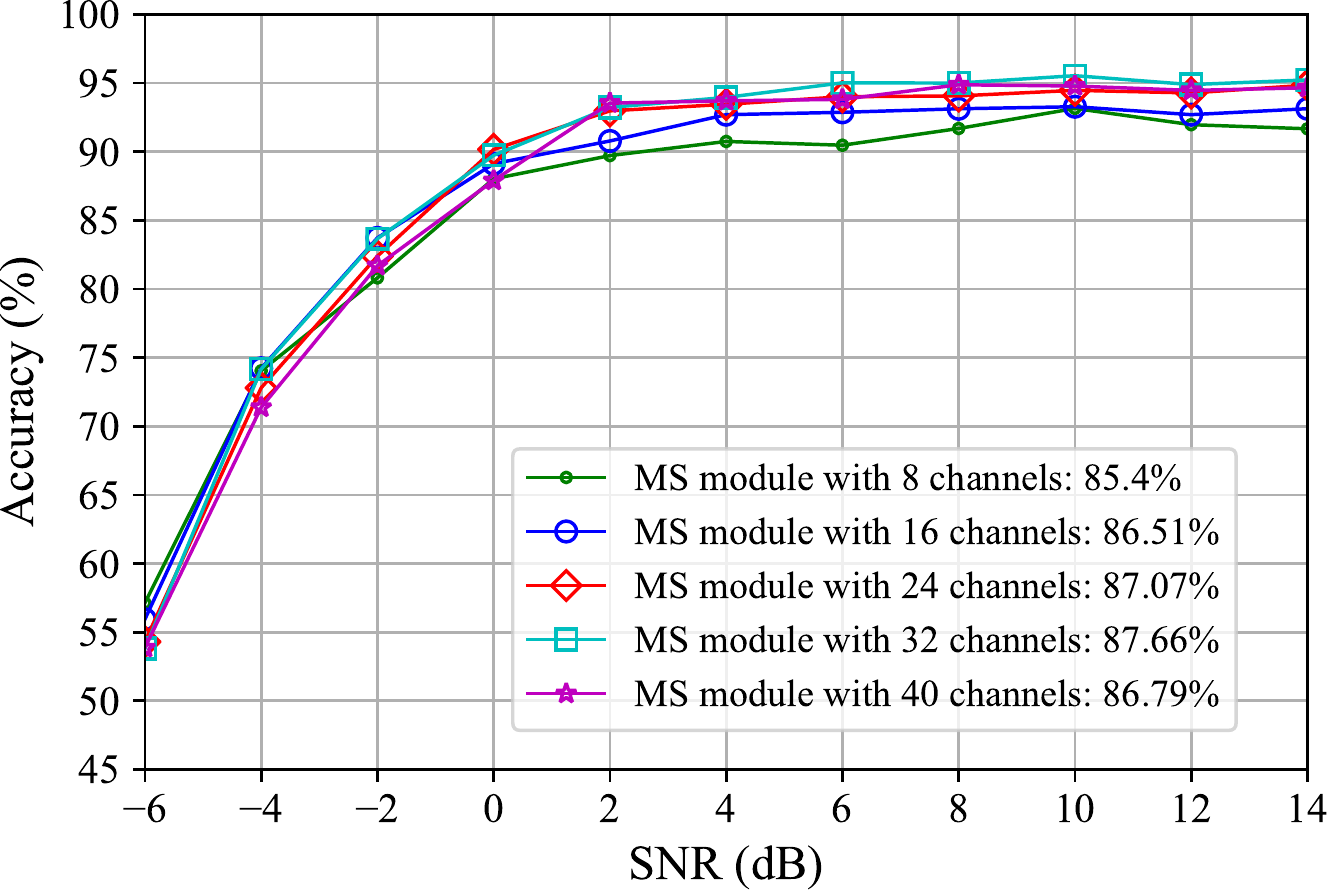}
\label{fig:channel}
\end{minipage}
}
\centering
\caption{
Performance comparison of (a) MSNet with different MS modules and (b) MS module with different channels.
}
\end{figure*}

It has been demonstrated that the depth of the network has an influence on the performance of the deep learning models\cite{simonyan2014very}. Thus, an evaluation on the depth of the proposed MSNet is performed. According to \cite{OShea2018}, a fixed channel number of $32$ is used for different layers of the MSNet. 
Fig. \ref{fig:depth} shows the performance achieved with different numbers of multi-scale modules used in the MSNet, and the corresponding overall classification accuracy (the average accuracy achieved under the whole testing dataset) is given.
It is quite evident that the MSNet with two layers of the multi-scale modules has the best performance among all the schemes with an overall accuracy of $87.66\%$. It can also be observed that the increase of the number of MS modules does not improve the performance of the model but increases the computational cost of the network. Thus, two layers of the multi-scale modules are utilized in our proposed MSNet to achieve the best classification performance in the following experiments. Fig. \ref{fig:channel} shows the classification accuracy of the proposed multi-scale module with different numbers of channels. The accuracy of the MSNet first increases with the increase of the channel number of the multi-scale module. However, the performance becomes worse when the channel number reaches $40$ mainly due to the over-fitting problem. MS module with $32$ channels obtains the best performance. Specifically, when the number of the channels is larger than $16$ and the SNR is above $2$dB, the classification accuracy of the MSNet can be higher than $90\%$.

\begin{figure}[h]
	\includegraphics[width=0.95\linewidth]{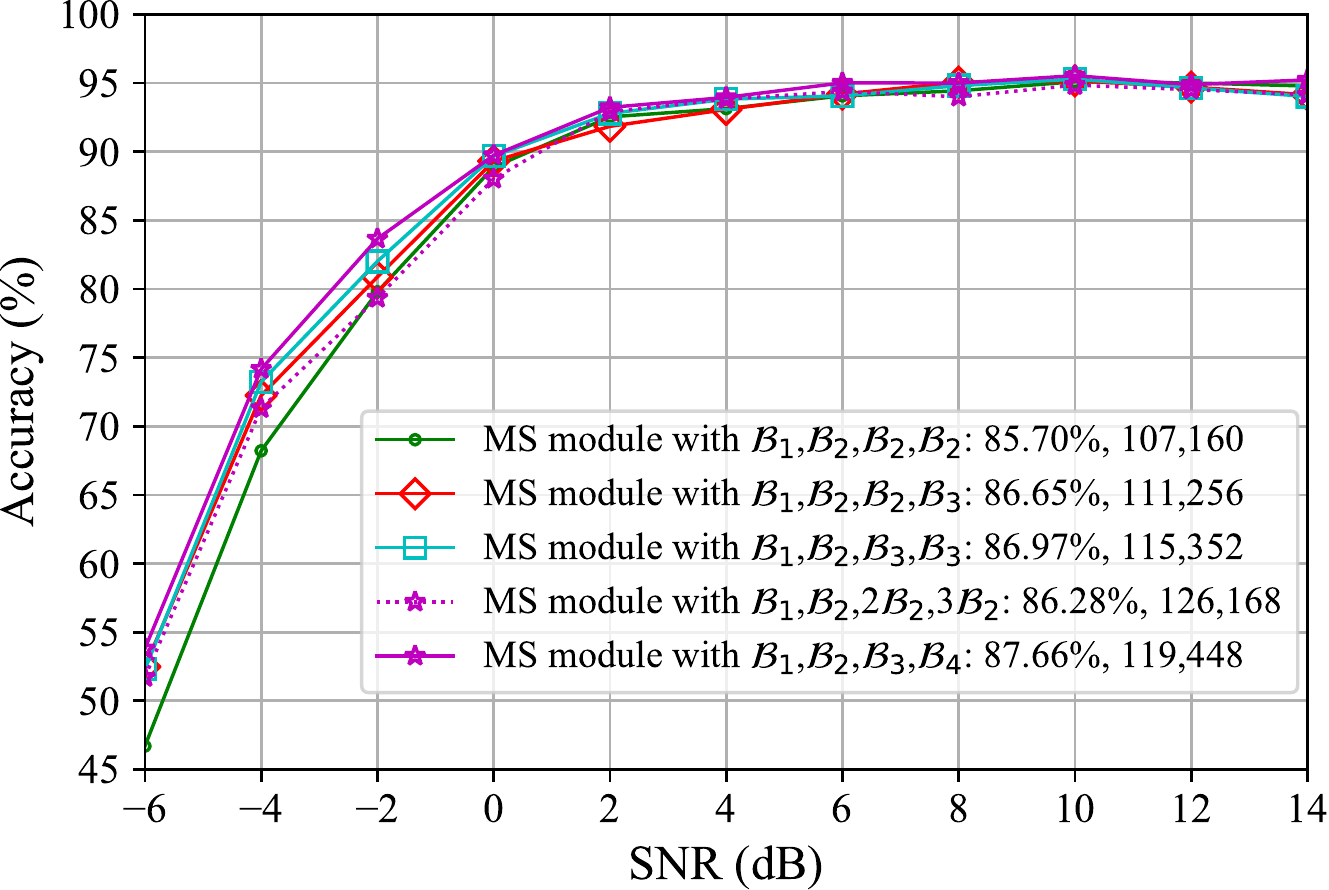}
	\caption{Classification performance of our proposed MS module with different kernel sizes.}
	\label{fig:kernel}
\end{figure}

\begin{figure}[h]
	\includegraphics[width=0.99\linewidth]{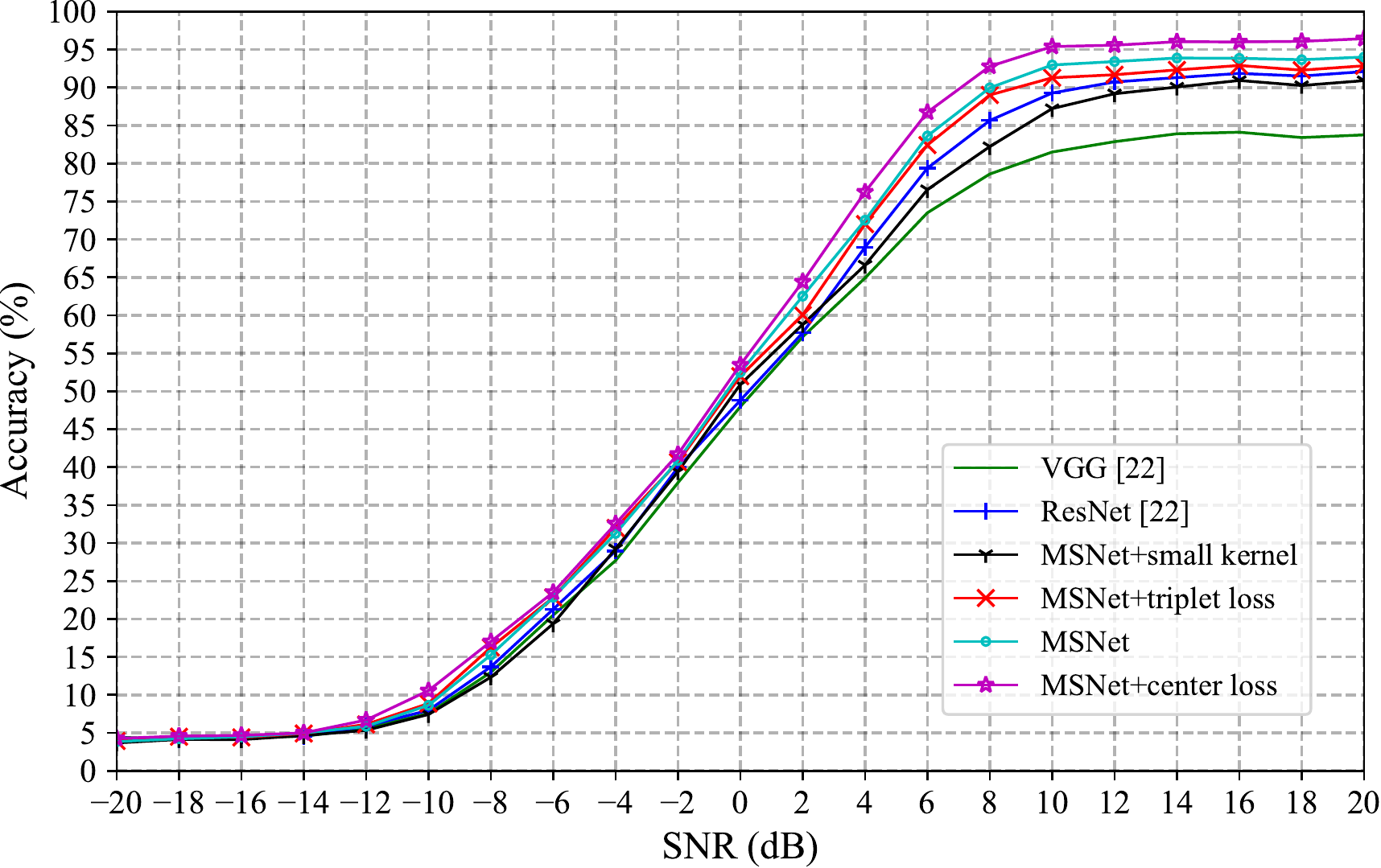}
	\caption{Classification performance comparison among MSNet, VGG\cite{OShea2018}, ResNet\cite{OShea2018}, MSNet with the triplet loss \cite{schroff2015facenet} and MSNet with small kernel under the RadioML 2018.01A \cite{OShea2018}
}
	\label{fig:comparison_18}
\end{figure}

Fig. \ref{fig:kernel} shows the classification accuracy of the proposed multi-scale module with different kernel sizes. Four branches of convolutional layers with different kernel sizes are compared, where $\mathcal{B}_i, i=\{1,2,3,4\}$, denotes $1\times1$, $3\times1$, $5\times1$, and $7\times1$, respectively. It can be seen that when replacing the smaller kernels with larger ones, the classification performance is increased. Moreover, to further show the performance  of large kernel sizes, a large kernel such as $\mathcal{B}_3=5\times1$ is replaced by a stack of two layers of $3\times1$ with an equal receptive field, and $\mathcal{B}_4=7\times1$ is equal to triple layers of $3\times1$ convolution. The accuracy of the MSNet with several equal size small kernels can only achieve $86.28$\%, which is about $1.5$\% lower than that of MSNet with large kernel sizes. Specifically, under low SNRs (when SNR is less than $0$dB), it is obvious that the multi-scale with larger kernels can achieve a better performance than the multi-scale with smaller kernels. 
Moreover, to evaluate the cost in a numerical manner, we calculate the network parameters under different kinds of kernels as shown in Fig. \ref{fig:kernel}, where $\mathcal{B}_1$, $\mathcal{B}_2$, $\mathcal{B}_3$, and $\mathcal{B}_4$ represent the conventional layers with size $1\times 1$, $3\times 1$, $5\times 1$, and $7\times 1$, respectively. It is evident that replacing the common $3\times 1$ kernels with large kernels of $5\times 1$ and $7\times 1$ only increases the network parameters from $107,160$ to $119,448$. 
Apart from the cost in computation, the boost in accuracy is evident. The accuracy improvement is about $2\%$ in the overall SNRs, and over $5\%$ accuracy gain is achieved under low SNR conditions. 
To further demonstrate the function of the larger kernel, we train the MSNet with small kernels (MS module with $\mathcal{B}_1$, $\mathcal{B}_2$, $\mathcal{B}_2$, $\mathcal{B}_2$) under a more complicated dataset RadioML 2018.01A. As shown in Fig. \ref{fig:comparison_18}, it is observed that a larger kernel can achieve a better performance than small one. It achieves about $2-5\%$ boost under low SNRs and over $7\%$ under high SNRs. 
Compared to the use of the equal small kernels (the combination of $\mathcal{B}_1,\mathcal{B}_2,2\mathcal{B}_2,3\mathcal{B}_2$) with $126,168$ parameters, large kernels can achieve a better performance with less parameters. 
Thus, the use of larger kernels for improving the accuracy results in a higher computational cost, while the cost is worthy for its boost in accuracy especially under the complicated scenarios with a wide SNR range.

\begin{figure}[h]
	\includegraphics[width=0.95\linewidth]{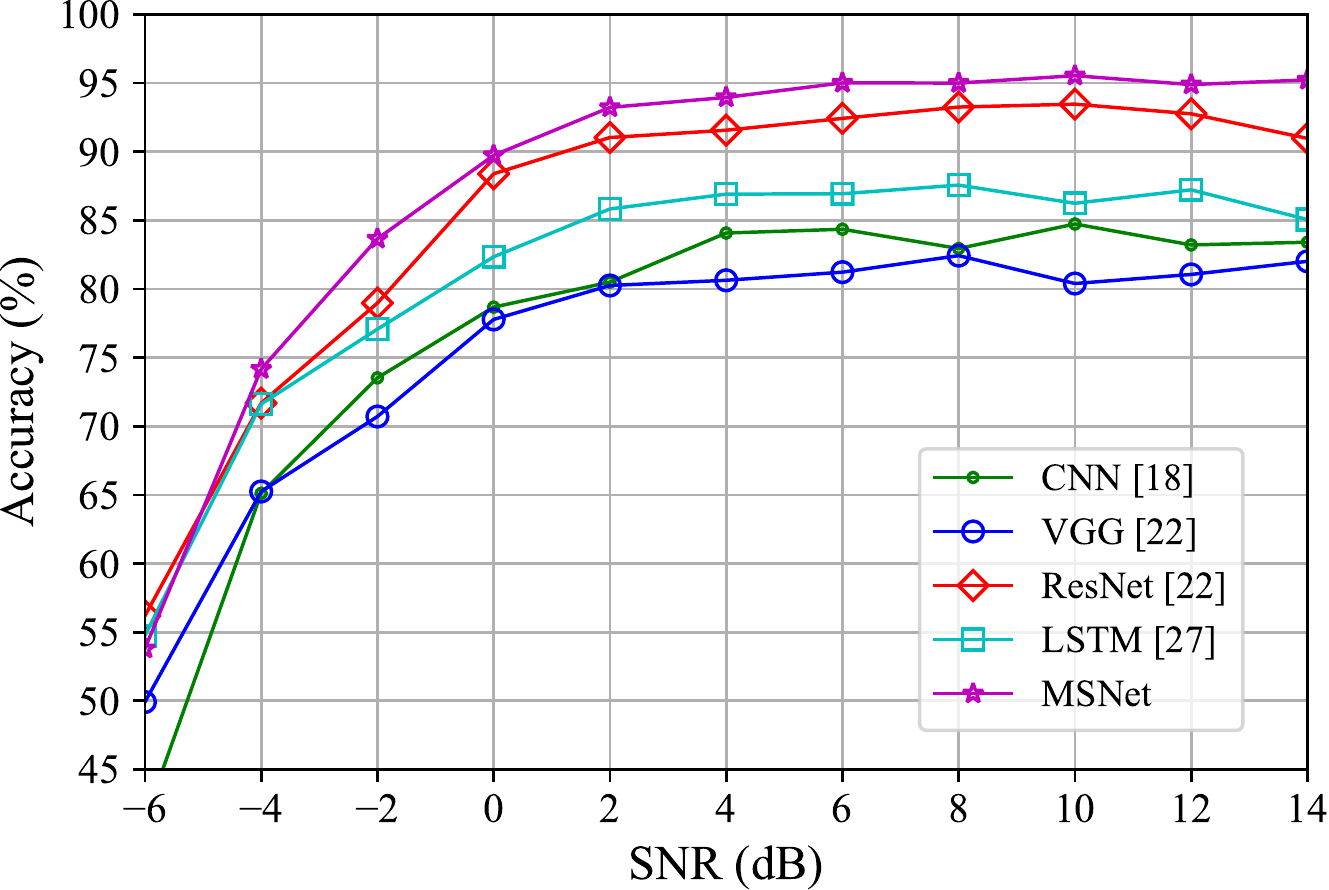}
	\caption{
	Classification performance comparison among MSNet, CNN\cite{OShea2016}, VGG\cite{OShea2018}, ResNet\cite{OShea2018} and LSTM\cite{rajendran2018deep} 
	}
	\label{fig:comparison}
\end{figure}

\begin{figure*}[h]
\centering
\subfigure[]{
\begin{minipage}[t]{0.45\linewidth}
\centering
	\includegraphics[width=0.95\linewidth]{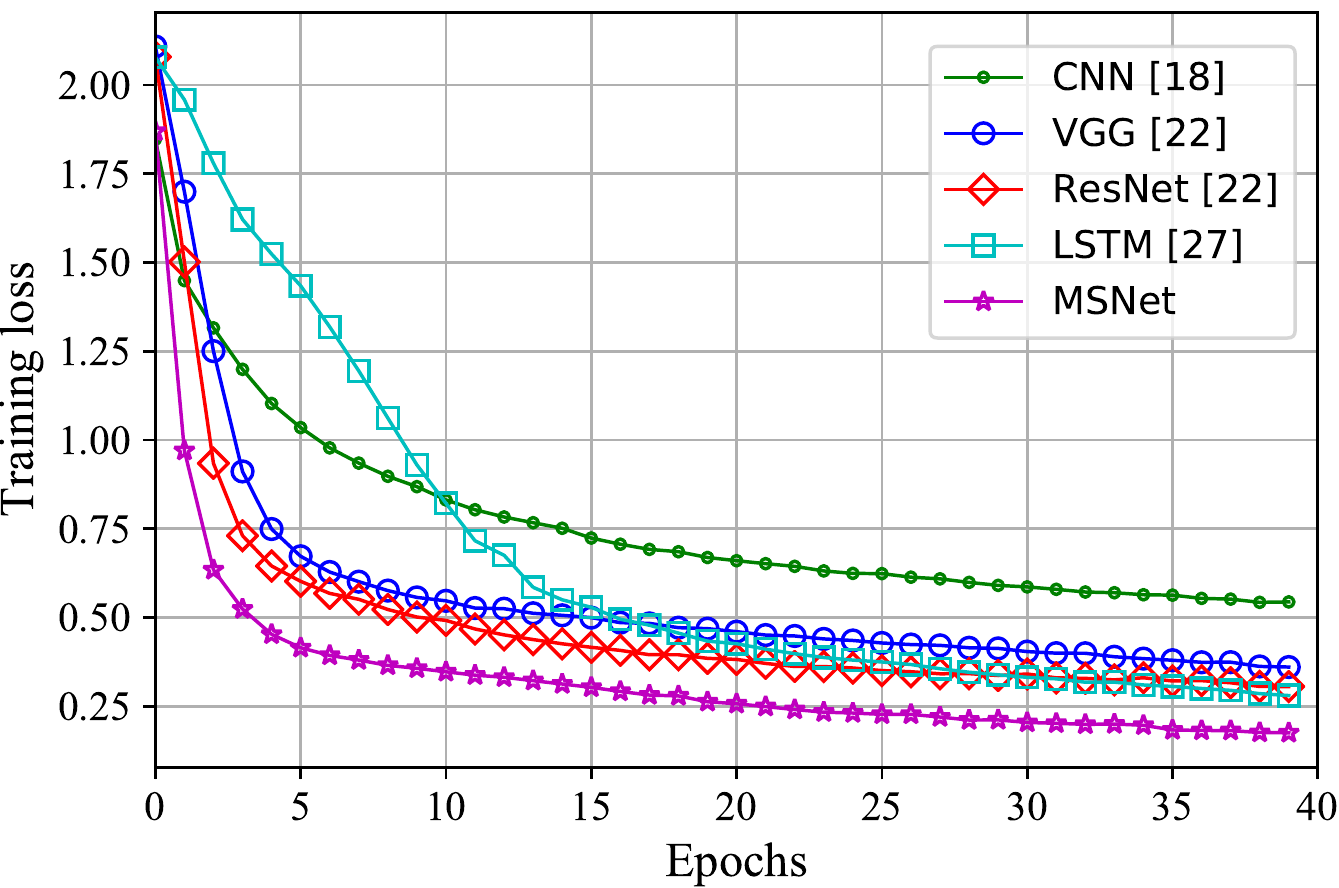}
\label{fig:train_loss}
\end{minipage}
}
\subfigure[]{
\begin{minipage}[t]{0.45\linewidth}
\centering
	\includegraphics[width=0.95\linewidth]{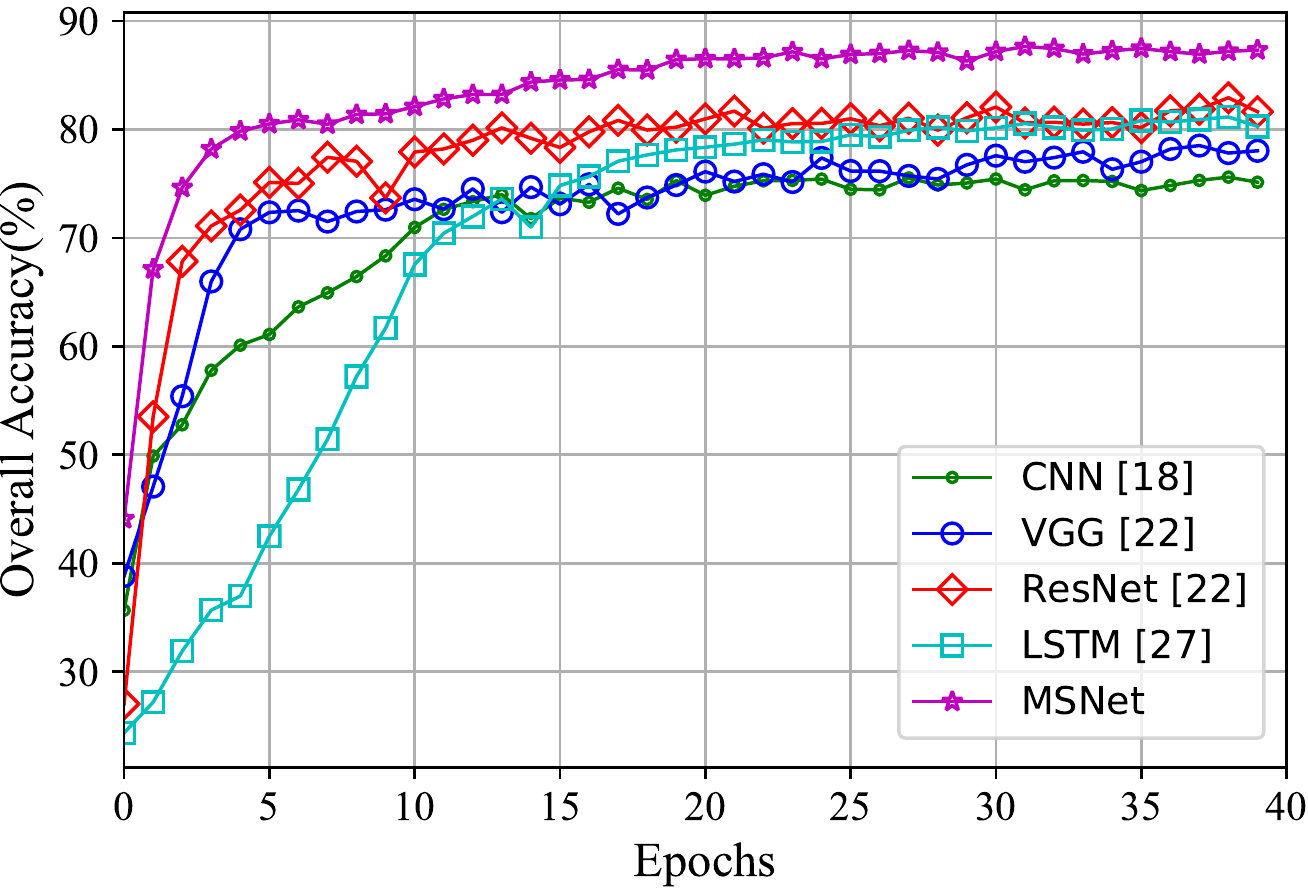}
\label{fig:test_loss}
\end{minipage}
}
\centering
\caption{Comparison of (a) training loss and (b) testing accuracy.}
\label{fig:loss}
\end{figure*}

To demonstrate the effectiveness of the proposed MSNet, we compare the performance of the proposed MSNet with those achieved with several representative DL-based models for AMC including CNN\cite{OShea2016}, VGG\cite{OShea2018}, ResNet\cite{OShea2018}, and LSTM\cite{rajendran2018deep}. The parameter settings of these models are based on the works in \cite{OShea2016,OShea2018} and \cite{rajendran2018deep}. Specifically, for VGG and ResNet, only three layers of the corresponding modules are used to fit the input length of $128$. 
The proposed MSNet is trained with the optimal hyper-parameters of two MS modules and $32$ channels.
Note that all the models are trained and tested with the same RadioML dataset. As shown in Fig. \ref{fig:comparison}, it is evident that the proposed MSNet is superior to other traditional schemes. MSNet can provide $2$ dB gain over ResNet and $4$ dB gain over LSTM. Moreover, MSNet can reach over $90$\% accuracy when the SNR is larger than $0$dB, and it can achieve over $95$\% accuracy at $6$dB, while the best performance of ResNet is $93.47$\% at $10$dB. To further show the superiority of the proposed MSNet, Fig. \ref{fig:train_loss} and \ref{fig:test_loss} illustrate the training loss and the testing accuracy during the process of each model, respectively. The proposed MSNet can achieve a lower loss in the training set compared to the other models, and obtains a higher classification accuracy than other schemes. Most importantly, compared to VGG\cite{OShea2018} and ResNet\cite{OShea2018} with a fixed convolutional kernel, the training of MSNet is much more stable mainly due to the multi-scale convolutional layers with larger kernels.

\begin{figure}
	\includegraphics[width=0.95\linewidth]{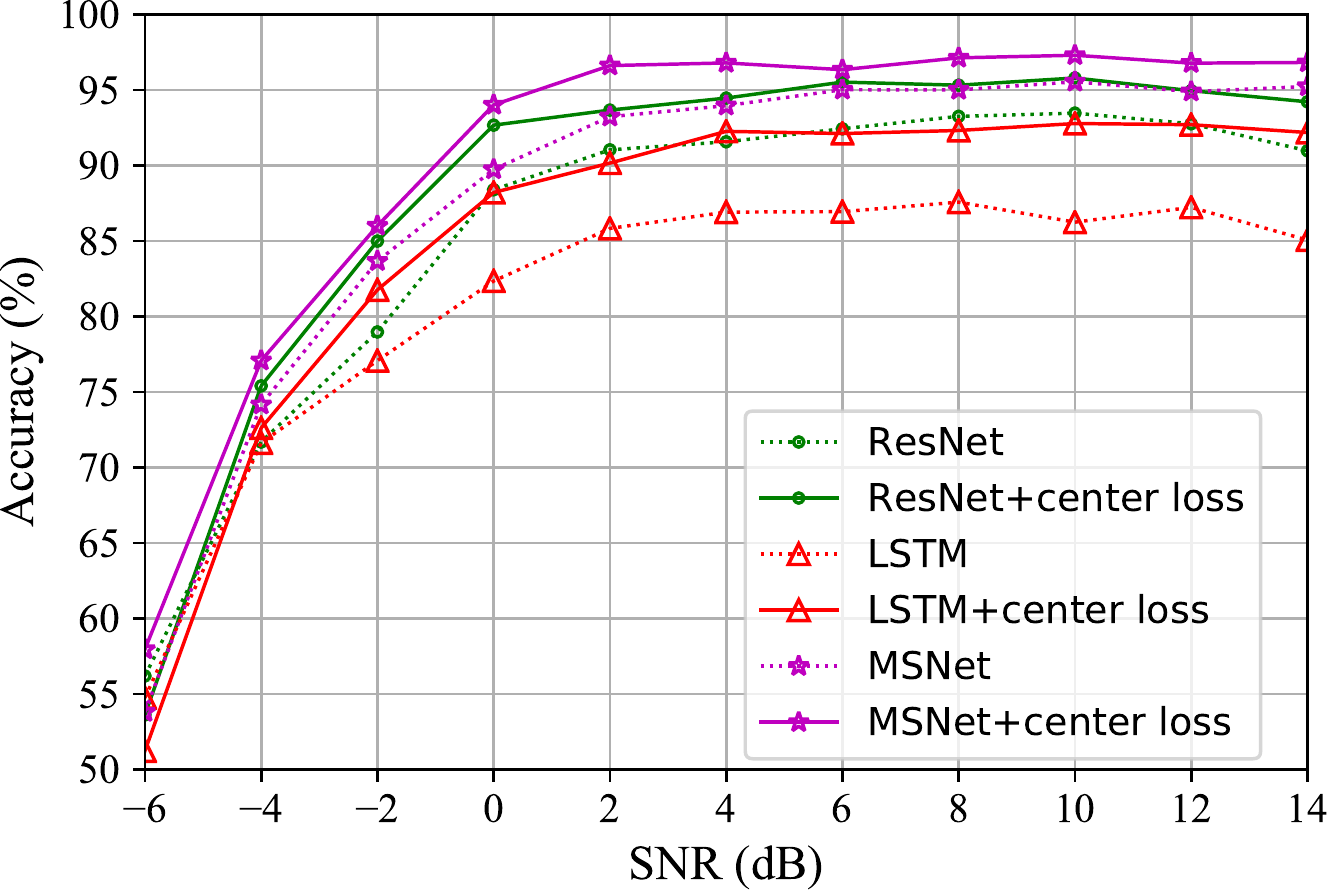}
	\caption{Classification performance comparison with models with and without center loss.}
	\label{fig:com}
\end{figure}

To learn the discriminative features for MSNet and other DL-based models, center loss is utilized during the training progress. To further demonstrate the effectiveness of the center loss and the two-stage training scheme, we operate the other two models ResNet \cite{OShea2018} and LSTM \cite{rajendran2018deep} with the center loss. As shown in Fig. \ref{fig:com}, the compared models can achieve about $2$ to $5$dB gain when the SNR is larger than -$2$dB. However, under the low SNR (less than -$2$dB) condition, the center loss does not help a lot and it even degrades  ResNet and LSTM while MSNet can still gain about $5$\% in accuracy. Specifically, with the assistance of center loss, MSNet can achieve over $95$\% classification accuracy when the SNR is larger than $1$dB, which is $5$dB more sensitive than the original MSNet. 

\begin{figure*}[h]
	\includegraphics[width=0.99\linewidth]{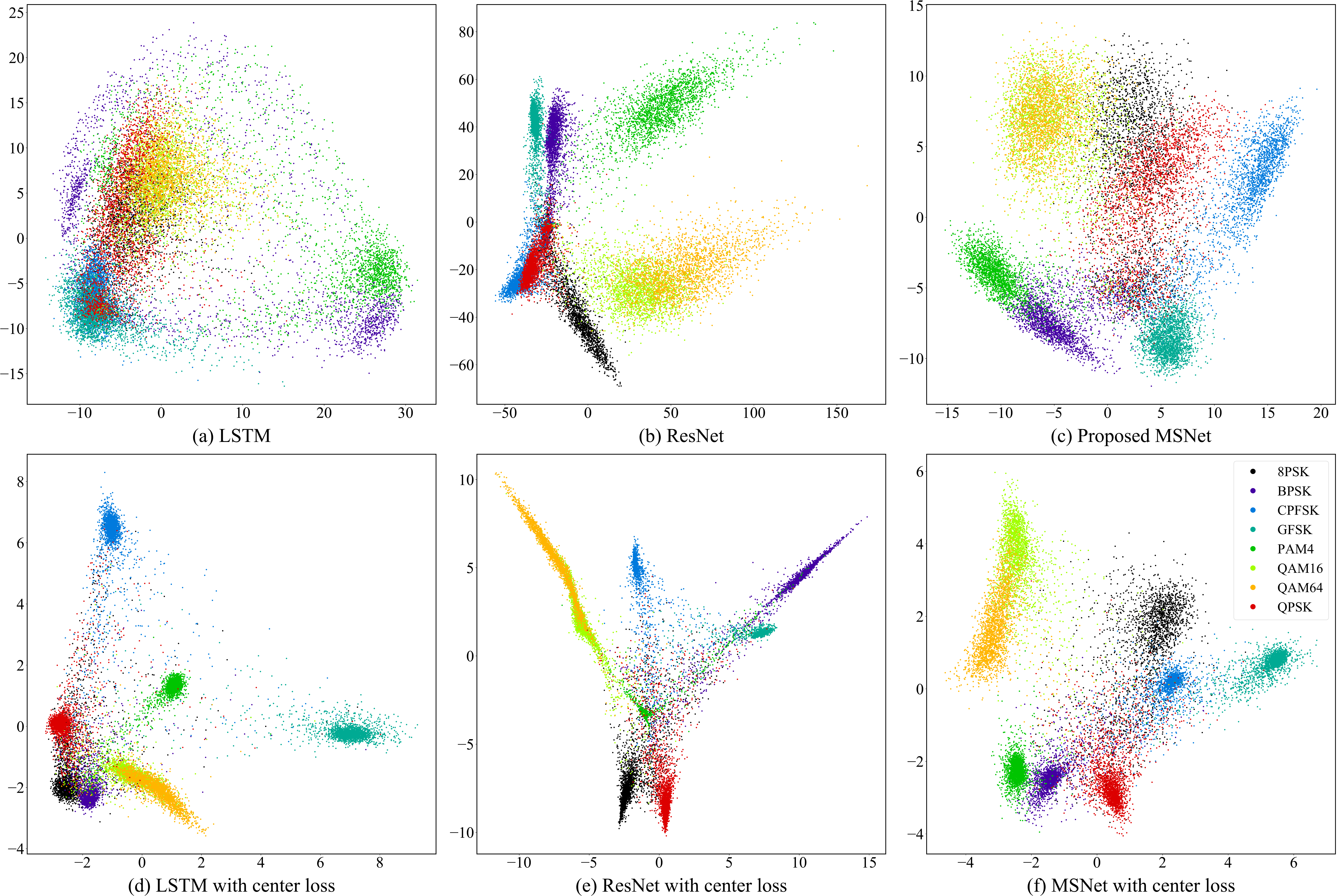}
	\caption{ Visualization scatter map of extracted features by different networks and center loss (after the PCA\cite{wold1987principal} operation).
}
	\label{fig:embeddings}
\end{figure*}

\begin{figure}
\centering
\subfigure[MSNet]{
\begin{minipage}[t]{0.45\linewidth}
\centering
\includegraphics[width=0.99\linewidth]{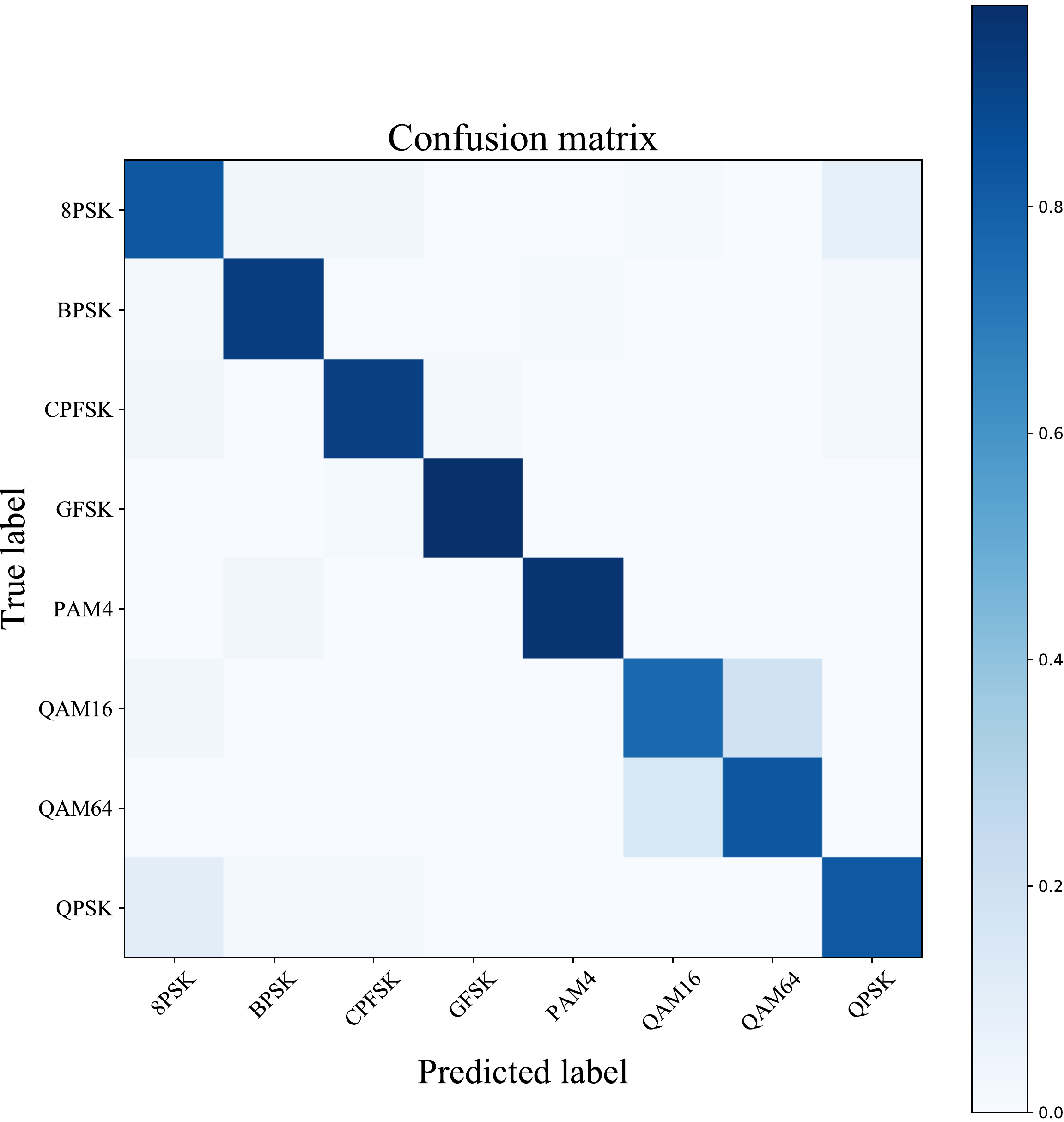}
\label{fig:msnetc}
\end{minipage}
}
\subfigure[MSNet with center loss]{
\begin{minipage}[t]{0.45\linewidth}
\centering
\includegraphics[width=0.99\linewidth]{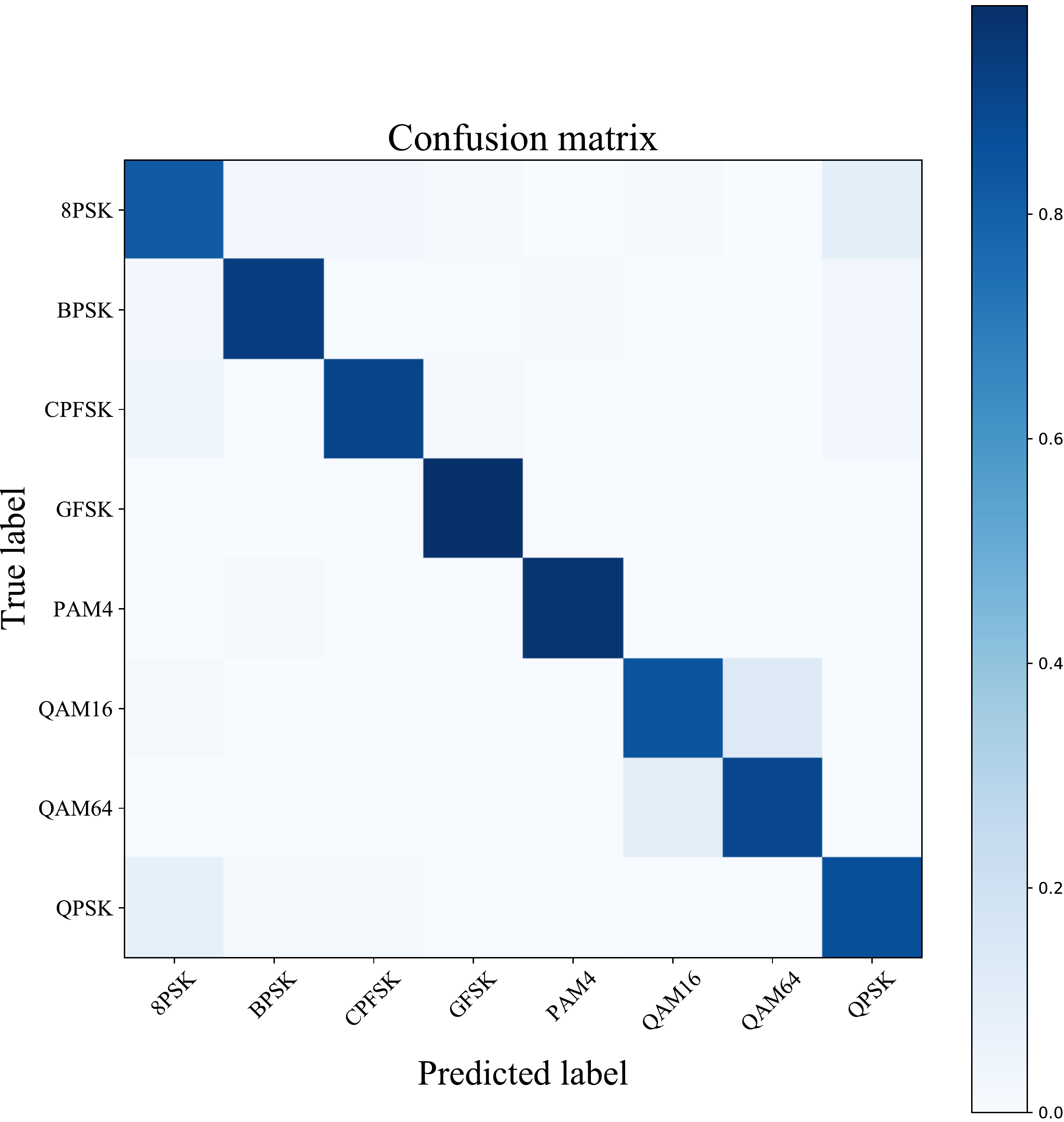}
\label{fig:msnetc_m}
\end{minipage}
}
\caption{Confusion matrix of classes of MSNet and MSNet with center loss.
}\label{fig:confusion}
\end{figure}

To understand the validity of the center loss in an intuitive manner, we visualize the features from the middle FC layer and convert them into a two-dimensional scatter map using the principal components analysis (PCA) \cite{wold1987principal} shown in Fig. \ref{fig:embeddings}. As shown in the top row, it is quite evident  that the features of the proposed MSNet are much more separable than those in LSTM \cite{rajendran2018deep} and ResNet \cite{OShea2018}. Additionally, it can also be seen that in these modulation formats, QAM$16$ and QAM$64$ are mixed together in the distribution view, which means that these modulations are very difficult to be classified. As illustrated in the bottom row of Fig. \ref{fig:embeddings}, with the novel loss function and the two-stage training strategy, these features are more clustered,  which means these features are much more discriminative. Specifically, for LSTM \cite{rajendran2018deep} in Fig. \ref{fig:embeddings}(d), the features are discriminative, while there still  exists overlap between these complex classes. ResNet \cite{OShea2018} mitigates this phenomenon to some extent, but the mis-classification of QAM$16$ and QAM$64$ is still worse, as  shown in Fig. \ref{fig:embeddings}(e). Compared to the other two models, the features extracted by our MSNet are much more discriminative and separable, as presented in Fig. \ref{fig:embeddings}(f). Moreover, the mixed features from QAM$16$ and QAM$64$ are much more separable. This can be further demonstrated by the confusion matrix in the class view. Fig. \ref{fig:msnetc} and \ref{fig:msnetc_m} show the class confusion matrix of the proposed MSNet and MSNet with center loss, respectively, where the horizontal axis denotes the predicted labels and vertical axis represents for the true labels. It is clearly seen that the mis-classification between QAM$16$ and QAM$64$ is alleviated with the help of the learning  the discriminative features. Specifically, the overall classification accuracy of QAM$16$ obtains a gain of $8.31$\% under the joint supervision.

To demonstrate the effectiveness of the two-stage training strategy, Table. \ref{tab1} illustrates the overall classification performance of VGG \cite{OShea2018}, LSTM\cite{rajendran2018deep}, ResNet\cite{OShea2018}, and MSNet. The first row shows the models without the two-step training, which means that these models are trained with the joint supervision only, and in the second row these models are optimized further with the second stage training (with the \emph{softmax} loss). These models achieve a gain of $2.62$\%, which demonstrates the obvious superiority under the optimized results of the first stage training. It can be seen that the well performed models such as ResNet and our proposed MSNet only achieve less than $1$\% boost in accuracy. This is because the \emph{softmax} loss in the second training step (\textbf{S2}) mainly focuses on learning separable features. For the well-designed models,  they are more capable to  learn separable features (as MSNet shown in Fig. \ref{fig:embeddings}(c)), and then, during the first training step (\textbf{S1}) with the joint supervision, the features are already both separable and discriminative. Consequently, the two-stage training strategy may not help much for the well-designed models with a high accuracy.
Moreover, for the complexity of the training processing, the two-stage training can be treated as a process of training (the first stage) and fine-tuning (the second stage). 
In the first stage, a constant learning rate is utilized to learn both discriminative and separable features by using the combined loss of the center loss and the \emph{softmax} loss. 
Then, in the second stage, the \emph{softmax} loss is used to further improve the performance. 
The second stage of the training process often overs within $5$ epochs, which is equal to the required epochs in the process of reducing the learning rate and fine-tuning the trained model. 
Thus, the two-stage training strategy cannot increase the complexity of the training. 

\begin{table}[h]
\centering
\caption{Overall Accuracy Comparison of Two-step Training Strategy}
\begin{center}
\begin{tabular}{ccccc}
\toprule
Models & VGG\cite{OShea2018}&  LSTM\cite{rajendran2018deep} & ResNet\cite{OShea2018} & MSNet \\
\midrule
w/o two-step training & 81.00\%  & 82.99\% & 87.35\% & \textbf{89.75\%} \\
\midrule
w/ two-step training  & 83.62\%  & 85.3\% & 88.24\% & \textbf{90.26\%} \\
\midrule
gains  & \textbf{2.62\%} & 2.31\% & 0.89\% & 0.51\% \\
\bottomrule
\end{tabular}
\label{tab1}
\end{center}
\end{table}

To further demonstrate the effectiveness of our proposed MSNet, 
we train the models in the large dataset RadioML 2018.01A \cite{OShea2018}. 
As shown in Fig. \ref{fig:comparison_18}, it is evident that the proposed MSNet can achieve the best classification performance than other benchmark schemes. 
The proposed MSNet with center loss can obtain about $3$ to $5$ dB gains compared to VGG \cite{OShea2018} and ResNet \cite{OShea2018}. 
Compared to using the \emph{softmax} loss, the center loss can help the MSNet achieve a gain of about $3\%$ in accuracy at above $0$dB. 
Even under low SNR conditions (SNR larger than $-10$dB), MSNet with the center loss can reach a boost of about $2\%$ compared to VGG and ResNet. 
Moreover, to validate the function of the center loss used for the proposed scheme, 
we also train the MSNet by replacing the center loss with the triplet loss in our scheme. 
It is evident that the triplet loss is not helpful for improving the accuracy, 
and even the performance becomes worse at high SNR conditions. 
That is mainly because the triplet loss can enlarge the intra-class distances, and the original \emph{softmax} has the same function. 
Thus, the combination of the triplet loss and the \emph{softmax} loss can damage the balance between the inter-class and intra-class distances. 
Compared to the triplet loss or the contrastive loss, the combination of the center loss and the \emph{softmax} loss is much more suitable for AMC.

\section{Conclusion}
A novel AMC scheme was proposed by using a novel loss function and MSNet in order to address the intra-class diversity problem confronted by the existing AMC schemes due to the dynamic changes of the wireless communication environment. Moreover, a novel loss function combining the center loss and \emph{softmax} loss was exploited to learn both discriminative and separable features. Simulation results demonstrated that our proposed scheme is superior to other benchmark schemes in terms of the classification accuracy. Moreover, the influences of the network parameters and the effectiveness of the novel loss function with the two-stage training strategy on the achievable performance were comprehensively investigated. 

\bibliographystyle{IEEEtran}
\bibliography{reference}

\end{document}